\documentclass[amsfonts,amsmath,prd,preprint,nofootinbib]{revtex4}

\usepackage{epsfig,bbm,cancel,ulem}
\usepackage[breaklinks=true]{hyperref}
\usepackage{xcolor}
\usepackage{wasysym}
\usepackage[mathscr]{eucal}
\usepackage{multirow}
\usepackage{amsmath}
\usepackage{enumitem}

\begin{document}

\title{Spherical orbits around Kerr-Newman and Ghosh black holes}

\author{A. S. Alam}
\email{agya.sewara@ui.ac.id}
\author{L. C. Andaru}
\email{langga.calvareno@ui.ac.id}
\author{B. N. Jayawiguna\footnote{Present Address: High Energy Physics Theory Group, Department of Physics, Faculty of Science, Chulalongkorn University, Bangkok 10330, Thailand.}}
\email{byon.nugraha91@ui.ac.id}
\author{H. S. Ramadhan\footnote{Corresponding author.}}
\email{hramad@sci.ui.ac.id}
\affiliation{Departemen Fisika, FMIPA, Universitas Indonesia, Depok, 16424, Indonesia.}
\def\changenote#1{\footnote{\bf #1}}

\begin{abstract}
We conduct a comprehensive study on spherical orbits around two types of black holes: Kerr-Newman black holes, which are charged, and Ghosh black holes, which are nonsingular. In this work, we consider both null and timelike cases of orbits. Utilizing the Mino formalism, all analytical solutions for the geodesics governing these orbits can be obtained.  It turns out that all spherical photon orbits outside the black hole horizons are unstable. In the extremal cases of both models, we obtain the {\it photon boomerangs}. The existence of charge in the Kerr-Newman allows the orbits to transition between retrograde and prograde motions, and its increase tends to force the orbits to be more equatorial. On the other hand, the Ghosh black hole, characterized by a regular core and a lack of horizons in certain conditions, presents the possibility of observable stable spherical orbits in the so-called {\it no-horizon} condition. As the Ghosh parameter 
$k$ increases, trajectories tend to exhibit larger latitudinal oscillation amplitudes. We observe that as the Ghosh parameter $k$ increases the trajectories tend to have larger latitudinal oscillation amplitudes. Finally, we investigate the existence of {\it innermost stable spherical orbits} (ISSOs). Both black holes demonstrate the appearance of two branches of ISSO radii as a function of the Carter constant 
$\mathcal{C}$. However, there are notable differences in their behavior: in the case of the Kerr-Newman black hole, the branches merge at a critical value, beyond which no ISSO exists, while for the Ghosh black hole, the transcendental nature of the metric function causes the branches to become complex at some finite distance.%We investigate these orbits, both null and timelike, and analyze their behaviour.
\end{abstract}

\maketitle
\thispagestyle{empty}
%\section{Introduction}
\setcounter{page}{1}

\section{Introduction}

It is well-known that any system possessing spherical symmetry and staticity will preserve its angular momentum, and as a result, the corresponding dynamics are confined on a plane. This is true in classical mechanics as well as in general relativity (GR). What is genuine in GR, among other things, is that not only massive particles but light can also orbit around a massive object. Perhaps the ultimate massive object in the universe is a black hole. Its visual image was first released in 2019 by the Event Horizon Telescope (EHT) collaboration~\cite{EventHorizonTelescope:2019dse} for M87*, and for Sgr A* three years later~\cite{EventHorizonTelescope:2022wkp}. For Schwarzschild we know that the photon circular orbit ({\it photon sphere}, $r_{ps}$) is $r_{ps}=3M$, while for Reissner-Nordstrom (RN) there are  $r_{ps}^{\pm}=3/2\left(M\pm\sqrt{M^2-8Q^2/9}\right)$. Both the Schwarzschild and RN unstable photon spheres are outside their corresponding shadow radii, thus can be observed. In reality, however, all astrophysical objects including black holes rotate. The appropriate metric solutions are the Kerr and the Kerr-Newman (KN)~\cite{Kerr:1963ud, Newman:1965my}. EHT observed the Sgr A* angular diameter $d_{sh} = 48.7\pm 7 \mu as$, which exhibits consistency with theoretical predictions of Kerr black hole~\cite{EventHorizonTelescope:2022wkp}. Moreover, the observation confirms that the Sgr A* shows the viability of the Kerr model with constrained charge (KN)~\cite{Ghosh:2022kit,Vagnozzi:2022moj, Uniyal:2022vdu}%,Khodadi:2022pqh, KumarWalia:2022aop, KumarWalia:2022ddq, Uniyal:2022vdu}. 

The rotating metric introduces a departure from spherical symmetry, allowing for the existence of non-planar orbits, the {\it spherical orbits} (SO) being one of them. These orbits maintain a constant radius and trace the surface of a sphere. Their discovery stemmed from Carter's insight in 1968~\cite{Carter:1968rr}, where he identified that rotation introduces an additional Killing tensor, leading to the conservation of a quantity known as "{\it Carter's constant}" $C$ through Noether's theorem. Building upon this foundation, Wilkins in 1972 showcased the existence of "{\it spherical timelike orbits}" (STOs) in the extremal Kerr metric. Subsequently, Johnston and Ruffini extended these findings to the Kerr-Newman case~\cite{Johnston:1974pn}.

The exploration of "{\it spherical photon orbits}" (SPO) began with Stuchlik's investigation of the Kerr model~\cite{Stuchlik1981}, followed by Calvani, De Felice, and Nobili's study of the Kerr-Newman scenario~\cite{Calvani:1980is}. However, all studies on SPOs suggest that orbits are stable only within the inner horizon. Nonetheless, SPOs play a crucial role in the observation of black holes. It is the multiple unstable photon orbits around black holes that contribute to the formation of the luminous ring detected by instruments like the Event Horizon Telescope (EHT)~\cite{EventHorizonTelescope:2019dse, EventHorizonTelescope:2022wkp}. This discovery holds significant implications for our understanding of black hole phenomena.

%Rotating metric, however, lacks spherical symmetry. This opens up the possibility of non-planar orbits, the {\it spherical orbits} (SO), among other things. These are orbits with constant radii that trace the surface of a sphere. Their existence was realized when Carter~\cite{Carter:1968rr} pointed out that rotation introduces an additional Killing tensor to the system. By Noether's theorem, this new symmetry corresponds to a conserved quantity called the {\it Carter's constant}. Utilizing this conserved quantity, Wilkins~\cite{Wilkins:1972rs} demonstrated the existence of {\it spherical timelike orbit} (STO) in the extremal Kerr. Later, these results were generalized by Johnston and Ruffini to the Kerr-Newman case~\cite{Johnston:1974pn}. The study of {\it spherical photon orbits} (SPO) was initiated by Stuchlik for the Kerr model~\cite{Stuchlik1981}, and later by Calvani, De Felice, and Nobili for the Kerr-Newman case~\cite{Calvani:1980is}. On the other hand, all SPO studies suggest that the orbit is stable when it is inside the inner horizon. Therefore, the physical SPOs are all unstable. Moreover, SPO is extremely useful for the observation of black holes. In fact, it is the photon rays that orbit the black hole multiple times before escaping out due to instability that produce the ring of light detected by the EHT (Event Horizon Telescope)~\cite{EventHorizonTelescope:2019dse, EventHorizonTelescope:2022wkp}. 

To determine the trajectory, one must solve geodesic equations. The first numerical solutions were obtained by Goldstein~\cite{goldstein}. Mino demonstrated in~\cite{Mino:2003yg} that the geodesic equations can be decoupled, paving the way for Fujita and Hikida to achieve analytical solutions to the timelike geodesic equations in terms of elliptic integrals~\cite{Fujita:2009bp} (see also \cite{Battista:2022krl}). Further analytical investigations into Kerr(-Newman) geodesics can be found in works such as~\cite{Hackmann:2010zz, Hackmann:2013pva, Lammerzahl:2015qps, Wang:2022ouq}. Teo leveraged this formalism to derive a one-parameter class of analytical solutions for spherical photon orbits~\cite{teo} and timelike orbits~\cite{Teo:2020sey}. These exact spherical orbits have been extensively explored in subsequent studies (see~\cite{Grossman:2011ps, Tavlayan:2020cso, Tavlayan:2021ylq, Hod:2013mgr}).

%To find the trajectory, one must solve geodesic equations with constant radii. The first numerical solutions were obtained by Goldstein~\cite{goldstein}. Mino in~\cite{Mino:2003yg} shows that the geodesic equations can be decoupled. This later paved the way for Fujita and Hikida to solve the timelike geodesic equations analytically in terms of elliptic integrals~\cite{Fujita:2009bp}. Further studies on analytical solutions of the Kerr(-Newman) geodesics can be found, for example, in~\cite{Hackmann:2010zz, Hackmann:2013pva, Lammerzahl:2015qps, Wang:2022ouq}. By employing this formalism, Teo found a one-parameter class of analytical solutions for the spherical photon orbits~\cite{teo} and timelike orbits~\cite{Teo:2020sey}. These exact spherical orbits have since been extensively studied (see~\cite{Grossman:2011ps, Tavlayan:2020cso, Tavlayan:2021ylq, Hod:2013mgr}).  

Additionally for null trajectories, a family of parameters exists that causes the photon to return to the same point in the opposite direction after one azimuthal oscillation, satisfying the condition $\Delta\phi=\pi$. This phenomenon, termed the "photon boomerang," was introduced by Page~\cite{Page:2021rhx}. In the case of Kerr black holes, it occurs in the near-extremal case, $a\approx 0.994 M$, with zero angular momentum. Page demonstrated that the photon boomerang does not occur in naked singularities (NS) for pure Kerr. However, in~\cite{Anjum:2023axh}, it was shown that a Kerr black hole surrounded by {\it perfect fluid dark matter} (PFDM) enables the existence of the photon boomerang under NS conditions. Related works on PFDM can be found in~\cite{Das:2020yxw, Atamurotov:2021hoq, Atamurotov:2021hck, Das:2021otl, Narzilloev:2020qtd}.

%For null trajectories there also exists a family of parameters that makes the photon return to the same point in the opposite direction with the condition ($\Delta\phi=\pi$) after one azimuthal oscillation. Page called it the ``photon \textit{boomerang}"~\cite{Page:2021rhx}. For Kerr, it occurs at the near-extreme parameter ($a\approx 0.994 M$) with zero angular momentum. For pure Kerr, Page showed that the photon boomerang does not appear in naked singularity (NS). In \cite{Anjum:2023axh}, the authors show that the Kerr black hole surrounded by {\it perfect fluid dark matter} (PFDM) (see \cite{Das:2020yxw,Atamurotov:2021hoq,Atamurotov:2021hck,Hou:2018avu,Haroon:2018ryd,Das:2021otl,Narzilloev:2020qtd}) enables the existence of photon boomerang in NS condition. 

%\textcolor{red}{Moreover, the EHT observation (and related literature \cite{Ghosh:2022kit,Vagnozzi:2022moj,Khodadi:2022pqh,KumarWalia:2022aop,KumarWalia:2022ddq,Uniyal:2022vdu}) confirms that the black hole observation in Sgr A* shows the viability of the Kerr model with the constraint charges (KN).}

Singularity stands as the most generic aspects of black holes. It is, however, widely believed that the singularity might be resolved within the framework of quantum gravity. The concept of a non-singular black hole was initially proposed by Sakharov~\cite{Sakharov:1966aja} and later realized by Bardeen~\cite{Bardeen}. In the Bardeen metric, the near-core region exhibits behavior similar to the de Sitter space, and all invariants remain nonsingular throughout. Ayon-Beato and Garcia (ABG) demonstrated that the Bardeen metric can be derived from coupling Einstein's equations with a nonlinear electrodynamics (NLED) source~\cite{Ayon-Beato:1998hmi, Ayon-Beato:2000mjt}. For an in-depth review of regular black holes, readers are referred to~\cite{Ansoldi:2008jw} and the accompanying references.

Regular black holes were subsequently extended to the rotating case by Bambi and Modesto~\cite{Bambi:2013ufa}, as well as by Toshmatov et al.~\cite{Toshmatov:2014nya} using the Newman-Janis algorithm. The resulting metrics are characterized by singularity-free Petrov type D solutions, although they suffer from violations of the weak energy condition. Ghosh~\cite{Ghosh:2014pba} introduced a probability distribution function to mitigate these violations, resulting in a class of three-parameter stationary, symmetric metrics describing regular (nonsingular) rotating black holes. These metrics depend on the mass  ($M$), spin ($a$), and a free parameter ($k$) that quantifies the deviation from Kerr and generalizes the Kerr-Newman solution. Extensive literature discussing these solutions and their implications, including constraints from observations by the Event Horizon Telescope (EHT), can be found in~\cite{Amir:2016cen, Kumar:2020yem, Kumar:2020cve, Kumar:2020ltt}.

This paper aims to extensively explore the spherical orbit dynamics around Kerr-Newman and Ghosh black holes. In Section~\ref{sec:KN}, we delve into the behavior of spherical orbits around the Kerr-Newman (KN) black hole. We provide exhaustive analytical solutions and explicitly describe trajectories of the corresponding spherical photon orbits (SPOs) in Section~\ref{sec:SPOKN}, and spherical timelike orbits (STOs) in Section~\ref{sec:STOKN}. Transitioning to Section~\ref{sec:regu}, we investigate the spherical orbit solutions of the (regular) Ghosh black hole. There, we present analytical solutions and conduct a thorough analysis of various types of SPOs and STOs trajectories in Section~\ref{sec:SPOR} and Section~\ref{sec:STOR}, respectively. Finally, in Section~\ref{sec:conc}, we synthesize and summarize our findings.

%This paper aims to extensively study spherical orbit behaviour around Kerr-Newman and Ghosh black holes. In Sec.~\ref{sec:KN} we discuss spherical orbits around the Kerr-Newman (KN) black hole. We present complete analytical solutions as well as explicit trajectories of the corresponding SPO in Sec.~\ref{sec:SPOKN} and Spherical Timelike Orbits (STO) in Sec.~\ref{sec:STOKN}. In Sec.~\ref{sec:regu} we study the spherical orbits solutions of (regular) Ghosh black hole. The analytical solutions and complete analysis of various types of SPO and STO trajectories are discussed in Sec.~\ref{sec:SPOR} and Sec.~\ref{sec:STOR}, respectively. Finally in Sec.~\ref{sec:conc}, we summarize our findings.
%It is well known there are light orbits around Kerr black holes that are not necessarily confined in the equatorial plane, but light can orbit in non-equatorial orbits with constant radii \cite{teo}. These orbits are known as \textit{spherical photon orbits}. Spherical photon orbits mark the threshold between non-equatorial orbits that plunge into the black hole and those that do not. Such orbits play an important role in modeling the capture of light by black holes.

%\newpage

\section{Spherical Orbits around the Kerr-Newman Black Hole}
\label{sec:KN}

%\subsection{Equation of Motion}

The Kerr-Newman (KN) solutions in Boyer-Lindquist coordinate are~\cite{Newman:1965my}
\begin{equation}
\label{KNmetric}
    ds^2 = -\frac{\Delta}{\Sigma}\left(dt - a \sin^2\theta d\phi\right)^2 + \frac{\sin^2\theta}{\Sigma}\left[\left(r^2 + a^2\right)d\phi - a dt\right]^2 + \Sigma \left(\frac{dr^2}{\Delta} + d\theta^2\right),
\end{equation}
and
\begin{equation}
A_{\mu}=\frac{Qr}{\Sigma}\bigg(1,0,0,-a\sin^2\theta\bigg),
\end{equation}
where
\begin{eqnarray}
%\begin{split}
    \Sigma&\equiv& r^2 + a^2 \cos^2{\theta},\nonumber\\
    \Delta&\equiv& r^2 + a^2 - 2 M r + Q^2, \\
%\end{split}
\label{KNSigma}
\end{eqnarray}
and $M$, $Q$, $a$ denote the black hole’s mass, charge, and angular momentum per unit mass, respectively. The allowed range of 
$a$ is
\begin{equation}
0 \leq a^2 \leq M^2 - Q^2,
\end{equation}
where the upper and lower bounds correspond to the extremal Kerr-Newman and Reissner-Nordström (RN) black holes, respectively. The event horizons are $r_{\pm}\equiv M \pm \sqrt{M^2 - a^2 - Q^2}.$
%We are interested in investigating the orbit outside the outer horizon, the range of $r$ is between $r_+<r<\infty$. The range of ${t,\theta,\phi}$ takes the usual ranges.}
By defining $u\equiv\cos\theta$ the geodesic equations can be derived using the Hamilton-Jacobi formalism \cite{Carter:1968rr, Johnston:1974pn, Wang:2022ouq}:
\begin{eqnarray}
\label{KNgeod1}
\Sigma \Dot{t} & = & \left(\frac{r^2+a^2}{\Delta}\right) P(r) - a \left[ a E (1-u^{2}) - L_{z}\right],\\
\label{KNgeod2}
\Sigma \Dot{r} & = & \sqrt{R(r)},\\%\label{KNgeod2}\\
\label{KNgeod3}
\Sigma \Dot{u} & = & \sqrt{\Theta(u)},\\
\label{KNgeod4}
\Sigma \Dot{\phi} & = & \frac{a}{\Delta}P - \left(aE - \frac{L_z}{1-u^{2}} \right),
%\label{KNgeod4}
\end{eqnarray}
with
\begin{eqnarray}
\label{extraKN}
R(r) &\equiv& P(r)-\Delta[\mathcal{C}+\mu^2r^2+(L_{z}-a E)^2 ],\nonumber\\
\Theta(u) &\equiv&\mathcal{C}-\left[\mathcal{C}+L_z^2+a^{2} (\mu^2-E^2)\right]u^2+a^{2}(\mu^2-E^2)u^4,\\
P(r)&\equiv&\left[E(r^2 + a^2)-a L_{z}\right]^2.\nonumber
\end{eqnarray}
The photon (/timelike) case appears when $\mu=0$ $(/\mu=1)$. The dot refers to the derivative wrt the Affine parameter $\tau$, $\mathcal{C}$ is the Carter's constant, $E$ and $L_{z}$ are the energy and angular momentum about $\phi$-axis of the photon, respectively.

%5\textcolor{blue}{\begin{eqnarray}
%\label{extraKN}
%R(r) &\equiv& P^2-\Delta[(L_z-aE)^2 +\mathcal{C}],\nonumber\\
%\Theta(u) &\equiv&\mathcal{C}-\left[\mathcal{C}+L_z^2-E^2a^2\right]u^2-a^2u^4,\\
%P&\equiv&E(r^2 + a^2)-a L_z,\nonumber
%\end{eqnarray}
%where the dot refers to the derivative with respect to the affine parameter $\tau$ along the geodesic, $\mathcal{C}$ is the Carter's constant, $E$ and $L_z$ are the energy and angular momentum about $\phi$-axis of the photon, respectively.} 

\subsection{Spherical Photon Orbits}
\label{sec:SPOKN}

The SPO conditions are Eqs.~\eqref{KNgeod1}-\eqref{extraKN} with $\mu=0$.
%\begin{eqnarray}
%\label{extraKNphoton}
%R(r) &\equiv& P^2-\Delta[\mathcal{C}+(L_{z}-a E)^2 ],\\
%\Theta(u) &\equiv&\mathcal{C}-u^2 \left[E a^2-\frac{L_{z}^2}{1-u^2}\right],\\
%P&\equiv& E(r^2 + a^2)-a L_{z}.
%\end{eqnarray}}
The solutions for Eq.~\eqref{KNgeod3} are physical whenever $\Theta(u)\geq0$. The bounds of $u$ values are the roots of $\Theta(u_0)=0$. Since $\Theta(u)$ is not $Q-$dependent, the roots are exactly the same as in the Kerr case~\cite{teo, Teo:2020sey}.
% The geodesic equations can be written as
%\begin{eqnarray}
%\label{KNgeod}
%\Sigma \Dot{t} & = & \left(\frac{r^2+a^2}{\Delta}\right) P-a(aE \sin^2\theta - L_z), \nonumber\\
%\Sigma \Dot{r} & = & \sqrt{R(r)}, \nonumber\\
%\Sigma \Dot{u} & = & \sqrt{\Theta(u)}, \\
%\Sigma \Dot{\phi} & = & \frac{a}{\Delta}P - \left(aE - \frac{L_z}{\sin^2 \theta} \right).\nonumber
%\end{eqnarray}
Defining
\begin{equation}
\Phi \equiv \frac{L_z}{E},\ \ \ \  C \equiv \frac{\mathcal{C}}{E^2},
\end{equation}
we have, for $C>0$,
\begin{equation}
\label{u0photon}
    u_0^2 = \frac{(a^2 - C - \Phi^2) + \sqrt{(a^2 - C - \Phi^2)^2 + 4 a^2 C}}{2a^2}.
\end{equation}
When $C<0$, the following condition must be satisfied
\begin{equation}
    a^2 - C - \Phi^2 > 0.
    \label{reqKNphoton}
\end{equation}

%This condition is also restrictive and would rule out a photon orbit case.}

The spherical orbit condition requires the corresponding radius ($\equiv r_{SO}$) satisfies
\begin{equation}
\label{socond}
R(r=r_{SO})=\frac{dR(r)}{dr}\bigg|_{r=r_{SO}}=0.
\end{equation}
Solving the two equations simultaneously, we get two conditions:
\begin{eqnarray}
(i) ~~\Phi &=&\frac{r_{SO}^2 + a^2}{a},\ \ \ \ \ \ \ \ \ \ C=-\frac{r_{SO}^4}{a^2};\label{rsolphotonKN1}\\
\nonumber\\
(ii) ~~\Phi&=& -\frac{a^2 M+a^2 r_{SO}-3 M r_{SO}^2+2 Q^2 r_{SO}+r_{SO}^3}{a (r_{SO}-M)}, \nonumber\\ 
C&=& -\frac{r_{SO}^2 \left[4 a^2 \left(Q^2-M r_{SO}\right)+\left(r_{SO} (r_{SO}-3 M)+2 Q^2\right)^2\right]}{a^2 (r_{SO}-M)^2}.
\label{rsolphotonKN2}
\end{eqnarray}
Note that both solutions~\eqref{rsolphotonKN1}-\eqref{rsolphotonKN2} satisfy
\begin{equation}
\label{socond2}
\frac{d^2 R(r)}{dr^2}\bigg|_{r=r_{SO}}>0,
\end{equation}
implying that they are unstable under radial perturbation~\cite{Wilkins:1972rs}.
The first condition~\eqref{rsolphotonKN1} is unphysical~\cite{teo} since it does not satisfy condition~\eqref{reqKNphoton}. In the second condition~\eqref{rsolphotonKN2} $C$ must take a positive sign since $C<0$ violates the constraint~\eqref{reqKNphoton}.% Thus, the only physically-allowed solution is when $C>0$.

%\subsection{Properties Spherical Photon Orbit around Kerr-Newman Black Hole}
%In this section, we are going to study the properties of spherical photon orbit. 

\begin{figure}%[h]
\centering
\begin{tabular}{cc}	\includegraphics[width=0.4\linewidth]{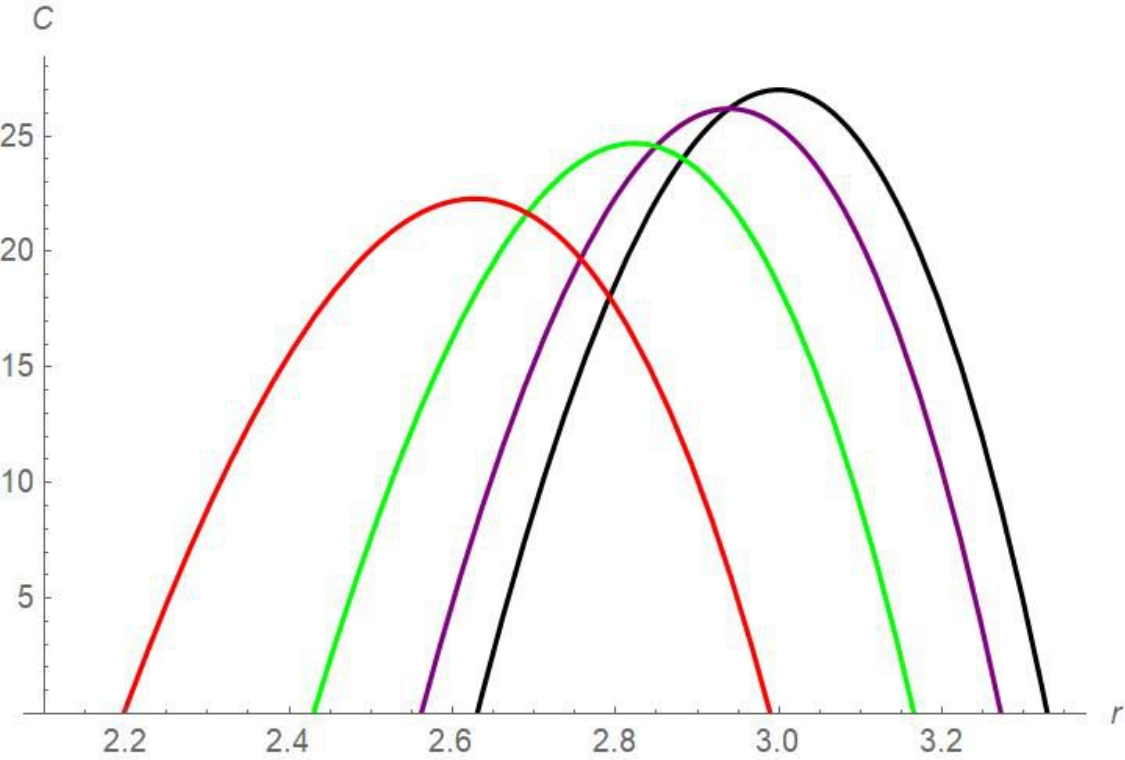} &
\includegraphics[width=0.4\linewidth]{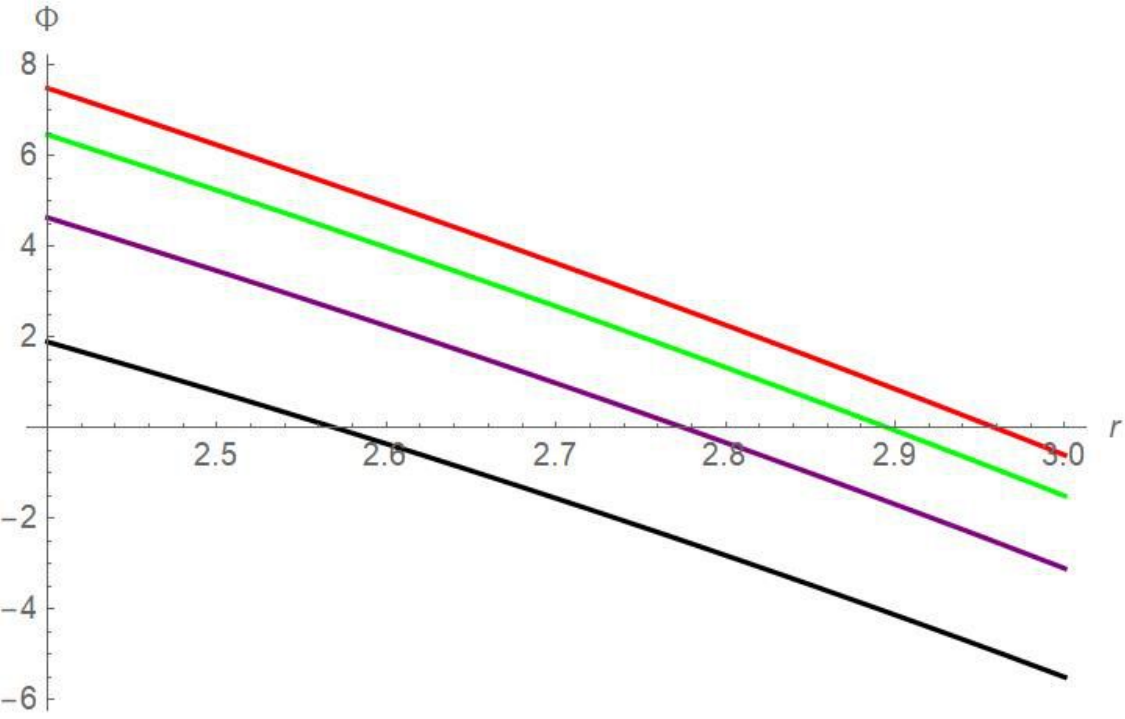}\\
\end{tabular}
\caption{Typical plots of $C$ and $\Phi$ as functions of $r_{SO}$ for several values of $Q$: $Q=0$ (black), $Q=0.3M$ (purple), $Q=0,5M$ (green) dan $Q=0,7M$ (red).}
\label{fig:Shift}
\end{figure}

Qualitatively, the plots of $C$ and $\Phi$ look similar to the Kerr case. However, they shift as $Q$ varies, as shown in Fig.~\ref{fig:Shift}.
%\begin{figure}
%\centering
%\includegraphics[scale=0.65]{PropKerr.pdf}
%\caption{The qualitative plot of $C$ (blue) and $\Phi$ (red) as functions of $r_{SO}$.}
%\label{fig:propkerr}
%\end{figure}
As $Q$ increases, the plot of $C$ shifts to the left while its maximum point decreases. Since Carter's constant $C$ represents the angular velocity of the photon in the equatorial direction (as in the Kerr case), for highly-charged black holes the orbits are forced to concentrate on the equatorial plane only.

%While the figure on the right of \ref{fig:Shift} is a plot of the shift in the angular momentum of the photon for the same charge value in the adjacent plot, in contrast to the smaller $C$ plot with an increasing charge value, the $\Phi$ plot will shift to the right. The orbit that is formed will be concentrated around the $z$ axis which can be interpreted as the formed orbit being concentrated in the equatorial plane.

\subsubsection{\bf Analytical Solutions}

The geodesic Eqs.~\eqref{KNgeod1}-\eqref{extraKN} can be solved analytically. Defining the Mino's parameter $\lambda$~\cite{Mino:2003yg} 
\begin{equation}
\frac{d\tau}{d\lambda}\equiv\Sigma\ \ \ \rightarrow\ \ \ \ \Sigma \frac{dx^\mu}{d\tau} = \frac{dx^\mu}{d\lambda}, 
\label{mino}
\end{equation}
the coordinates of $(\theta,\phi,t)$ can be solved in terms of the elliptic integral and the Jacobi elliptic functions \cite{Grad}. The Eq.~\eqref{KNgeod3} can be cast, by defining $w \equiv u^2$, as
\begin{equation}
    \frac{dw}{d\lambda} = \pm 2 Y(w),
    \label{lambdaKNPhoton}
\end{equation}
where
\begin{equation}
\label{yw}
Y(w)^2\equiv-a^2 w (w-w_1)(w-w_2),
\end{equation}
and $w_{1,2}$ are the roots of $Y(w)^2$:
\begin{equation}
    w_{1,2}= \frac{a^2-C-\Phi ^2 \pm \sqrt{4 a^2 C \left(-C-\Phi ^2+1\right)^2}}{2 a^2}.
\end{equation}
Choosing the positive sign and taking the initial condition $w(\lambda=0)=0$, Eq.~\eqref{lambdaKNPhoton} can be integrated
\begin{equation}
\lambda = \frac{1}{2} \int_0^w \frac{dw}{Y(w)} = \frac{1}{a\sqrt{w_1-w_2}}\ F(\Psi,k),
\label{lambdahasilphoton}
\end{equation}
where
\begin{equation}
\label{psik}
\Psi \equiv \arcsin{\sqrt{\frac{w(w_1-w_2)}{w_1(w-w_2)}}},\ \ \ 
k \equiv\sqrt{\frac{w_1}{w_1-w_2}},
\end{equation}
and $F(\Psi,k)$ is the {\it incomplete elliptic integral function of the first kind}. From Eq.~\eqref{lambdahasilphoton} it can be seen that $\lambda$ increases along with $w$. The one full latitudinal oscillation for $\lambda$ is
\begin{equation}
    \Delta \lambda = \frac{4}{a\sqrt{w_1-w_2}}K(k),
\end{equation}
with $K(k)$ is the {\it complete elliptic integral function of the first kind}.

To obtain the solution for $\theta$, it is necessary to write the inverse of Eq.~\eqref{lambdahasilphoton} in the form of the {\it Jacobi amplitude function}:
\begin{equation}
\label{amjacobi}
\Psi = \mathrm{am}(a\sqrt{w_1-w_2}\lambda,k).
\end{equation}
by writing the Eq.~\eqref{amjacobi} above in terms of the {\it elliptic sine} $sn$ function, 
\begin{equation}
\sin{\Psi} = \mathrm{sn}(a\sqrt{w_1-w_2}\lambda,k),
\end{equation}
the analytical solution of $\theta$ can be written as
\begin{equation}
\label{thetasol}
\theta(\lambda) = \arccos\left[{ \sqrt{-w_2}\ k\  \text{sd}\left(a \sqrt{w_1-w_2} \lambda,k \right)}\right],
\end{equation}
where $sd(x,y)\equiv sn(x,y)/dn(x,y)$ and $dn(x,y)$ is the so-called {\it the delta amplitude function} defined as 
\begin{equation}
dn(x,y)\equiv\frac{d}{dx}\mathrm{am}(x,y).
\end{equation} 

From Eq.~\eqref{mino}, the geodesic equations for $\phi$ and $t$ can be written
\begin{eqnarray}
\frac{d \phi}{dw} &=& \frac{1}{2Y(w)} \left[ \frac{a(r^2 +a^2 - a \Phi) }{\Delta} - a + \frac{\Phi}{1-w} \right], \\
\frac{d t}{dw} &=&  \frac{1}{2Y(w)} \left[ \frac{(r^2+a^2)^2 -(r^2+a^2) a \Phi}{\Delta} + a \Phi -a^2(1-w). \right]
\end{eqnarray}
Assuming the initial condition $\phi(w=0)=t(w=0)=0$, the solutions are
\begin{eqnarray}
\phi(\lambda)&=& \frac{\Phi}{a(1-w_2)\sqrt{w_1-w_2}} [ F(\Psi,k) - w_2 \Pi(\Psi,k^2(1-w_2),k) ]\nonumber\\ %\nonumber
&&+ \left[ \frac{a(r^2 +a^2 - a \Phi) }{\Delta} - a \right] \lambda,\label{phisol} \\ 
t(\lambda)&=& - \frac{a}{\sqrt{w_1-w_2}} [ (1-w_2) F(\Psi,k) + w_2 \Pi(\Psi,k^2,k) ]\nonumber\\ %\nonumber
&&+ \left[ \frac{(r^2+a^2)^2 -(r^2+a^2) a \Phi}{\Delta} + a \Phi \right] \lambda,\label{tsol} 
%\label{phitsolpkn}
\end{eqnarray}
where $\Pi(\Psi,w_1,k)$ is {\it the incomplete elliptic integral function of the third kind}. The above two solutions are valid for all values of the $\lambda$ parameter. When $Q=0$ the solution reduces to the Teo's~\cite{teo}. The changes in $\phi$ and $t$ for a period of $\Delta \lambda$, i.e. one complete latitudinal oscillation, are
\begin{eqnarray}
\Delta \phi &=& \frac{4}{\sqrt{w_1-w_2}} \left[ \frac{\Phi}{a(1-w_1)} \Pi(-\frac{-w_1}{1 - w_1},k) + \left( \frac{r^2 +a^2 - a \Phi }{\Delta} - 1 \right) K(k) \right],\label{delphi} \\ 
\Delta t&=&- \frac{4}{\sqrt{w_1-w_2}} \bigg\{ -a[(1-w_2) K(k) + w_2 \Pi(k^2,k)]\label{delte} \nonumber\\ %\nonumber
&&+ \left[ \frac{(r^2+a^2)^2 -(r^2+a^2) a \Phi}{a \Delta} + \Phi \right]K(k)\bigg\}. 
\end{eqnarray}
Eq.~\eqref{delphi} determines the type of spherical orbits formed around the black hole. When $\Delta\phi$ is positive the orbit is {\it prograde}, and when it is negative it becomes {\it retrograde}.

The analytical solutions~\eqref{thetasol} and~\eqref{phisol}-\eqref{tsol} completely specify the photon trajectories. In the following subsections, we shall present and discuss extensively their 3D plots based on the extremality of the black hole.

\subsubsection{\bf Selected Orbits}

In this subsection, we focus on the extremal Kerr-Newman (KN) black hole, as the behavior of orbits in the non-extremal case exhibits similar trends to those in the extremal case. The extremal KN black hole occurs when
\begin{equation}
a = \sqrt{M^2 - Q^2}.
\end{equation}
The constraints \eqref{rsolphotonKN1}-\eqref{rsolphotonKN2} can be used to determine the orbit parameters. Once both the angular momentum of the source, $a$, and the angular momentum of the test particle, $\Phi$, are fixed, the remaining parameters are completely determined.

\begin{figure}
\centering
\includegraphics[scale=0.6]{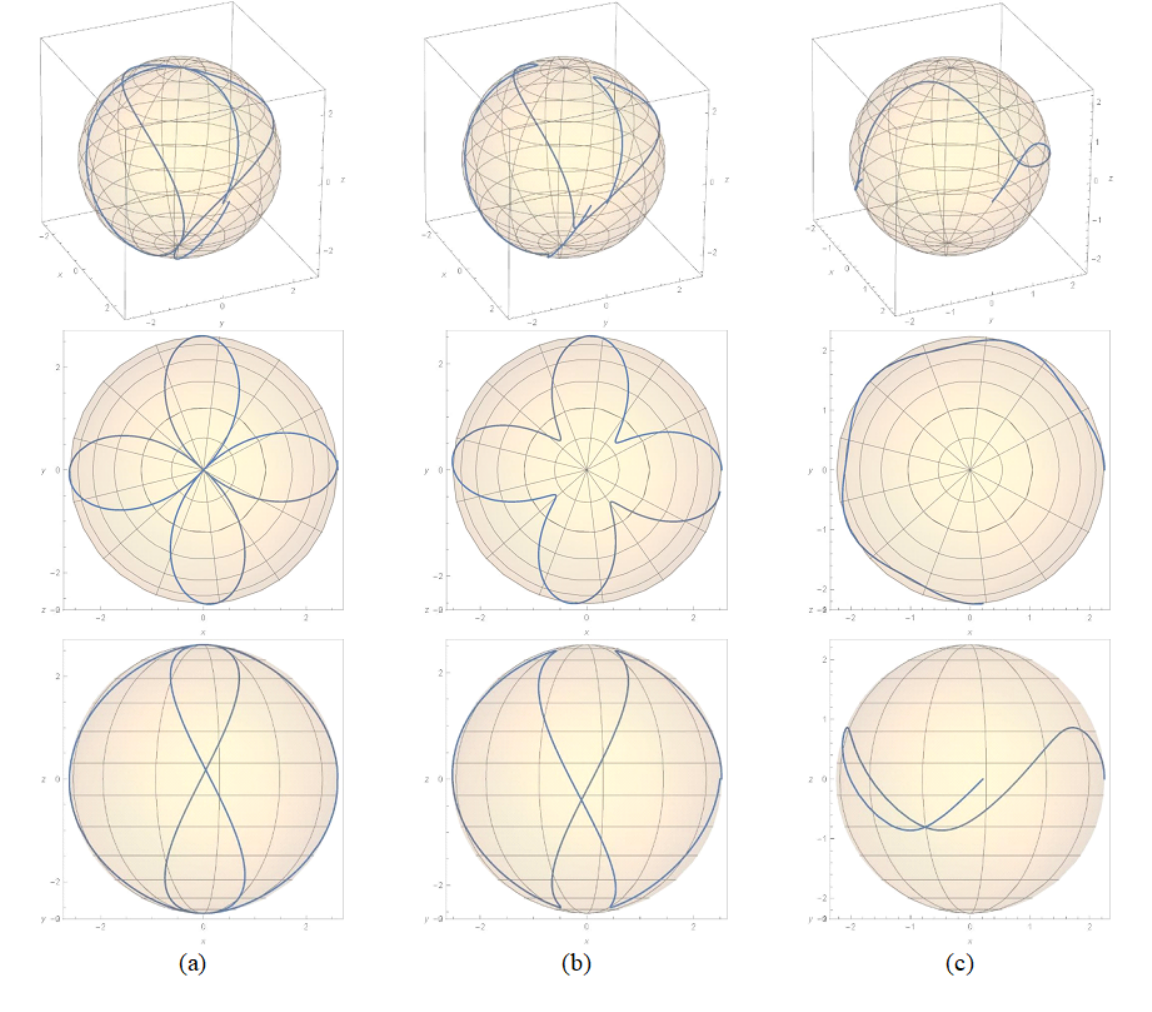}
\caption{Two latitudinal oscillations of the photon orbit around the extremal KN with $\Phi = 0$ in the $(x-y-z)$ as well as the projective $x-y$ and $x-z$ planes. The orbit begins from the equator and heads northwards. The direction of the black hole’s rotation is from west to east. Figures (a), (b), and (c) are cases with $Q=0$, $Q=0.4M$, and $Q=0.8M$, respectively.}
\label{fig:Phi0P}
\end{figure}

The orbit parameters for the extremal Kerr-Newman (KN) black hole are provided in Tables~\ref{tab3} and \ref{tab4} for $\Phi=0$ and $\Phi=-M$, respectively.
\begin{table}
\centering
\begin{tabular}{||c c c c c c||} 
 \hline\hline
\ \ Orbit\ \  &\ \ $Q/M$\ \ &\ \ $r/M$\ \ &\ \ $C/M^2$\ \ &\ \ $u_0$\ \ &\ \ $\Delta\phi$\ \ \\ [0.5ex] 
 \hline
(a) & 0 & 2.414 & 22.314 & 1 & 3.1761 \\ 
(b) & 0.4 & 2.356 & 21.372 & 1 & 2.473 \\
(c) &  0.8 & 2.166 & 18.410 & 1 & 0.596 \\
 \hline\hline
\end{tabular}
\caption{Parameters of the spherical photon orbit around extreme Kerr-Newman black hole when $\Phi=0$.}
\label{tab3}
\end{table}
In Fig.~\ref{fig:Phi0P}, we depict three examples of orbits for two latitudinal oscillations when the angular momentum of the photon $\Phi=0$ in the $(x-y-z)$ plane, the $x-y$ and $x-z$ planes respectively. Each plot showcases the orbit on an imaginary sphere with a fixed radius (details regarding the radius and other parameters can be found in Table~\ref{tab3}). The photons initiate their movement from the equator and progress northward in each orbit, mirroring the direction of rotation of the black hole itself, which is from west to east. The nine illustrations in Figs.~\ref{fig:Phi0P} (a), (b), and (c) correspond to cases where $Q=0$, $Q=0.4M$, and $Q= 0.8M$, respectively. Notably, the orbit depicted in (a) aligns with the one observed in Teo~\cite{teo}, where they touch one another at the poles. However, for $Q=0.4M$ the orbits, previously tangent to each other, begin to separate. The charge of the black hole begins to change the direction of the photons' orbits at the poles. Upon further increasing the charge to $Q\sim0.8M$, the two latitudinal oscillations are insufficient to complete an orbit. Additionally, the increased charge concentration renders the orbit more aligned with the equatorial plane, as evidenced in Fig.~\ref{fig:Phi0P} (c). This trend is attributed to the reduction in the black hole's angular momentum. Furthermore, we investigate the occurrence of a photon boomerang in this extremal case. The boomerang condition, 
\begin{equation}
    \Delta \phi = \pi,
\end{equation}
is satisfied when the charge value is small, $Q=0.02528M$ and $\Phi=0$. The corresponding plot closely resembles Fig.~\ref{fig:Phi0P} (a).

\begin{figure}%[H]
\centering
\includegraphics[scale=0.6]{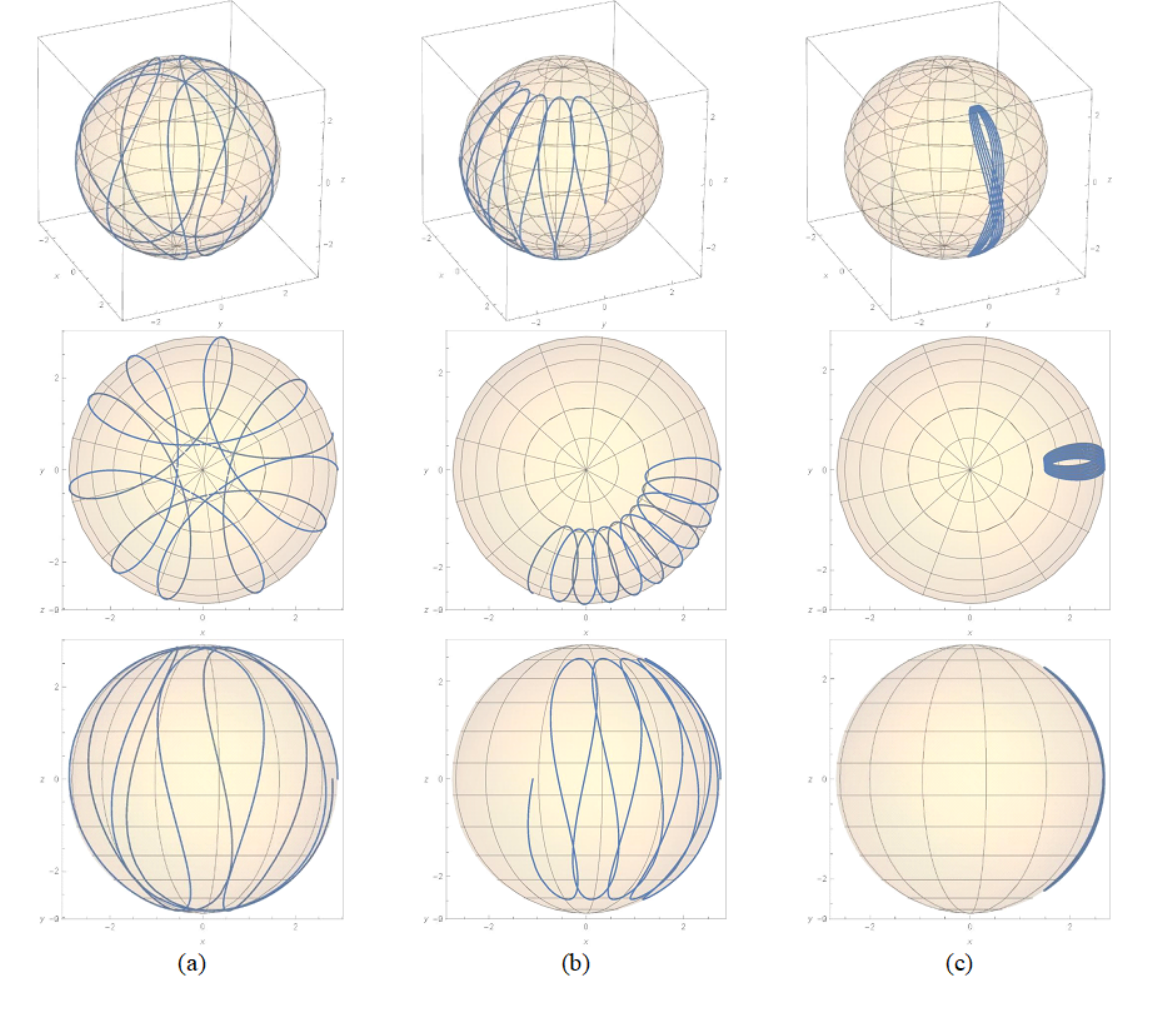}
\caption{Five latitudinal oscillations of the photon orbits around extremal KN with $\Phi = -M$ in the $(x-y-z)$ as well as the projective $x-y$ and $x-z$ planes. Figures (a), (b), and (c) are cases with $Q=0$, $Q=0.5M$, and $Q=0.6M$ respectively.}
\label{fig:Phi-1}
\end{figure}

In Fig.~\ref{fig:Phi-1} we illustrate three examples of SPO with $\Phi=-M$ for five latitude oscillations. Each orbit is plotted on an imaginary sphere with a fixed radius whose radius values, as well as other orbital parameters, can be found in Table~\ref{tab4}.
\begin{table}
\centering
\begin{tabular}{||c c c c c c||} 
 \hline\hline
\ \ Orbit\ \ &\ \ $Q/M$\ \ &\ \ $r/M$\ \ &\ \ $C/M^2$\ \ &\ \ $u_0$\ \ \ &\ \ \ $\Delta\phi$\ \ \ \\ [1ex] 
 \hline
\ \ (a)\ \ &\ \ 0\ \ &\ \ 2.732\ \ &\ \ 25.85\ \ &\ \ 1\ \ &\ \ -3.714\  \ \\ 
\ \ (b)\ \ &\ \ 0.5\ \ &\ \ 2.617\ \ &\ \ 23.92\ \ &\ \ 0.9803\ \ &\ \ -0.3959\ \ \\
\ \ (c)\ \ &\ \ 0.6\ \ &\ \ 2.562\ \ &\ \ 23.02\ \ &\ \ 0.9794\ \ &\ \ 0.0303\ \ \\ [1ex] 
 \hline\hline
\end{tabular}
\caption{Parameters of the spherical photon orbit around Kerr-Newman black hole when $\Phi=-M$.}
\label{tab4}
\end{table}
Figs.~\ref{fig:Phi-1} (a)-(c) depict the scenarios where  $Q=0$ (Kerr), $Q=0.5M$, and $Q=0.6M$,  respectively. The presence of charge alters the direction of the photon, and its augmentation prompts the orbit to return to the equatorial plane. Notably, for  $Q=0.6M$ an intriguing phenomenon happens: the photon undergoes a twist in its rotation direction about the $z$ axis. Initially retrograde ($Q<0.6M$), it abruptly transitions to prograde ($Q\geq0.6M$). The charge can effectively steer photons in the direction of the black hole's rotation, resulting in a ribbon-shaped orbit, as evidenced in Fig.~\ref{fig:Phi-1} (c).

%\newpage

%\section{Spherical Orbit around KN Black Hole}

\subsection{Spherical Timelike Orbits}
\label{sec:STOKN}

The timelike geodesics equations around a Kerr-Newman black hole remain consistent with Eqs.~\eqref{KNgeod1}-\eqref{extraKN}, with $\mu=1$. Bound and unbound orbits manifest when $E^2<1$ and $E^2>1$, respectively. Adhering to condition \eqref{socond}, typically yields four solutions. The initial two solutions are described by:
%\cite{kipp} are
%\begin{eqnarray}
%\Sigma \Dot{t} & = & \frac{r^2+a^2}{\Delta}P-a(aE \sin^2\theta - L_z), \\
%\Sigma \Dot{r} & = & \sqrt{R(r)}, \\
%\Sigma \Dot{u} & = & \sqrt{\Theta(u)}, \\
%Sigma \Dot{\phi} & = & \frac{a}{\Delta}P - \left(aE - \frac{L_z}{\sin^2 %\theta} \right),
%\end{eqnarray}
%\begin{eqnarray}
%R(r) &=& P^2 - \Delta[\mu^2 r^2 + (L_z - aE)^2 + c], \\ \nonumber
%\Theta(u) &=& c - \left[c + L_z^2 + (\mu^2 - E^2) a^2 \right]u^2 + a^2 %u^4(\mu^2 - E^2), \\ \nonumber
%P &=& E(r^2 + a^2) - a L_z. 
%\label{extramassive}
%\end{eqnarray}
%The $\mu$ notation is the rest mass of the particle, and for the timelike particle is $\mu=1$. 
%\subsection{Spherical Photon Orbit Requirement}
\begin{eqnarray}
\label{KNreq}
E_{a,b} & = & \sqrt{\frac{X \pm 2 a r_{SO}^2 Y \left(- M r_{SO}+Q^2\right) \left(a^2-2 M r_{SO}+Q^2+r_{SO}^2\right)}{\Gamma }}, \\ \nonumber
\Phi_{a,b} &=& \frac{a r_{SO}^2 Y \left(M r_{SO}-Q^2\right) \left[a^2+r_{SO} (r_{SO}-2 M)+Q^2\right]+\alpha }{\beta } E_{a,b},
\end{eqnarray}
where
\begin{eqnarray}
X &\equiv& r_{SO}^2 \bigg\{-2 a^4 \mathcal{C} \left(Q^2-M r\right)-a^2 s+r_{SO}^2 \left[r_{SO}(r_{SO}-3 M)+2 Q^2\right] \left[r (r_{SO}-2 M)+Q^2\right]^2\bigg\},\nonumber\\
\\
Y &\equiv& \sqrt{a^2 C^2+r_{SO}^2 \left[3 C M r_{SO}-r_{SO}^2 \left(C+Q^2\right)-2 C Q^2+M r_{SO}^3\right]}, \\
\Gamma &=& r_{SO}^6 \lbrace{ 4 a^2 (Q^2 - M r_{SO}) + [2 Q^2 + r_{SO} (-3 M + r_{SO})]^2 \rbrace}. \\
s &\equiv& \mathcal{C} \left[r_{SO}^2 \left(5 M^2-4 M r_{SO}+r_{SO}^2\right)+2 Q^2 r_{SO} (r_{SO}-3 M)+2 Q^4\right] \nonumber\\ 
&&+ r_{SO}^2 \left(M r-Q^2\right) \left[r_{SO} (3 r_{SO}-5 M)+2 Q^2\right], \\
\alpha &\equiv& a^2 r_{SO} \left(Q^2-M r_{SO}\right) \left[-M r_{SO}^2 \left(a^2+C+4 Q^2\right)+a^2 \mathcal{C} M+Q^2 r_{SO} \left(a^2+C+Q^2\right) \right] \nonumber\\
&&+ \left[ 2 r_{SO}^3 \left(2 M^2+Q^2\right)-3 M r_{SO}^4 \right] a^2 r_{SO} \left(Q^2-M r_{SO}\right), \\
\beta &\equiv& a\ r_{SO} \left(Q^2-M r_{SO}\right) \bigg\{ a^2 \left[C (M-r_{SO})+r_{SO} \left(Q^2-M r_{SO}\right)\right]\nonumber\\
&&+r_{SO} \left[r_{SO} (r_{SO}-2 M)+Q^2\right]^2\bigg\}.
\end{eqnarray}
The third and fourth solutions are related to the first two as the following
\begin{equation}
    (E_{c,d},\Phi_{c,d})=-(E_{a,b},\Phi_{a,b}),
\end{equation}
just like the Kerr case~\cite{Teo:2020sey}. The first two are, however, the only physical solution: $(E_{a,b},\Phi_{a,b})$.

\subsubsection{\bf Analytical Solutions}

The roots of $\Theta(u)=0$ (Eq.~\eqref{extraKN}) with $w\equiv u^2$ can be written as
\begin{equation}
    w_{1,2} = \frac{\left[a^2 \left(1-E^2\right)+\mathcal{C}+\Phi ^2 \mp \sqrt{\left[a^2 \left(1-E^2\right)+\mathcal{C}+\Phi ^2\right]^2-4 a^2 \mathcal{C} \left(1-E^2\right)} \right]}{2 \left(1-E^2\right)}.
\end{equation}
The ranges of $w_{1,2}$ depend on the sign of $E^2<1$ or $E^2>1$, and the sign of $C$. For bound orbit we consider $E^2<1$ and $0 \leq w \leq w_1 \leq 1 < w_2$. The exact solutions for $(\theta,\phi,t)$ are 
\begin{eqnarray}
\theta(\lambda) &=& \arccos\left( \sqrt{w_1}\ \text{sn}[a \sqrt{\left(1-E^2\right) w_2}\ \lambda,k] \right),\label{boundtimeKN1} \\
\phi(\lambda) &=& \frac{\Phi  }{a \sqrt{\left(1-E^2\right) w_2}} \Pi (\Psi,w_1,k) + \frac{a P-a \Delta  F}{\Delta }\lambda,\label{boundtimeKN2} \\
t(\lambda) &=& - \frac{a E}{\sqrt{(1-E^2)w_2}} [ (1-w_2) F(\Psi,k) + w_2 E(\Psi,k) ]\nonumber\\ 
&&+ \left[ \frac{(r^2+a^2)^2 -(r^2+a^2) a \Phi}{\Delta} + a \Phi \right] \lambda,\label{boundtimeKN3} 
\end{eqnarray}
where $E(\psi,k)$ is {\it the incomplete elliptic integral of the second kind}, and
\begin{equation}
\Psi\equiv\text{am}(a \sqrt{\left(1-E^2\right) w_2}\ \lambda,k),\ \ \ 
k\equiv\sqrt{\frac{w_1}{w_2}}.
\end{equation}
$\theta$ is a periodic function of $\lambda$, with period
\begin{equation}
    \Delta \lambda = \frac{4}{a \sqrt{\left(1-E^2\right) w_2}}K(k).
\end{equation}
The change in $\phi$ and $t$ for one period are
\begin{eqnarray}
    \Delta \phi &=& \frac{4}{a \sqrt{\left(1-E^2\right) w_2}} \left[ \frac{\Phi}{a} \Pi(w_1,k) + \frac{a(P-E\Delta)}{\Delta} K(k) \right], \\ 
    \Delta t &=& - \frac{4}{a \sqrt{\left(1-E^2\right) w_2}} \bigg\lbrace-a[(1-w_2) K(k) + w_2 E(k)] \nonumber\\
    &&+ \left[ \frac{(r^2+a^2)^2 -(r^2+a^2) a \Phi}{a \Delta} + \Phi \right]K(k)\bigg\rbrace. 
\end{eqnarray}

For $E^2>1$ the orbits become unbound, and the range of $w$ is $w_2 < 0 \leq w \leq w_1 \leq 1$. The exact solutions for $(\theta,\phi,t)$ are similar to the Eqs.~\eqref{boundtimeKN1}-\eqref{boundtimeKN3}, only that the elliptic functions have imaginary modulus~\cite{Teo:2020sey}. The case for $E^2=1$ is known as the {\it marginally bound orbits~\cite{Bardeen:1972fi}}.
%\begin{eqnarray}
%\theta(\lambda)&=& \arccos\left(\sqrt{-w_2}\ \text{sd}[a \sqrt{\left(E^2-1\right) (w_1-w_2)}\ \lambda,k]\right), \\
%\phi(\lambda)&=& \frac{\Phi}{a(1-w_2)\sqrt{(E^2-1)(w_1-w_2)}} [ F(\Psi,k) - w_2 \Pi(\Psi,k^2(1-w_2),k) ] \\ \nonumber
%    &+& \frac{a(P+\Delta E)}{\Delta}\lambda, \\ 
%    t &=& - \frac{aE}{\sqrt{(E^2-1)(w_1-w_2)}} [ (1-w_2) F(\Psi,k) + w_2 \Pi(\Psi,k^2,k) ] \\ \nonumber
%    &+& \left[ \frac{r^2+a^2}{\Delta}P + a \Phi \right] \lambda, 
%\end{eqnarray}
%where
%\begin{eqnarray}
%\Psi &\equiv& \text{am}(a \sqrt{\left(E^2-1\right) (w_1-w_2)}\lambda,k), \\ \nonumber
%k &\equiv& \sqrt{\frac{w_1}{w_1-w_2}}.
%\end{eqnarray}
%$\theta$ is a periodic function of $\lambda$, with period
%\begin{equation}
%    \Delta \lambda = \frac{4}{a \sqrt{\left(E^2-1\right) (w_1-w_2)}}K(k).
%\end{equation}
%The change in $\phi$ and $t$ for one period are 
%\begin{eqnarray}
%    \Delta \phi &=& \frac{4}{\sqrt{(E^2-1)(w_1-w_2)}} \left[ \frac{\Phi}{a(1-w_1)} \Pi(-\frac{-w_1}{1 - w_1},k) + \left( \frac{r^2 +a^2 - a \Phi }{\Delta} - E \right) K(k) \right], \\ 
%    \Delta t &=& - \frac{4}{\sqrt{(E^2-1)(w_1-w_2)}} \bigg\lbrace -aE[(1-w_2) K(k) + w_2 \Pi(k^2,k)] \\ \nonumber
%    &+& \left[ \frac{(r^2+a^2)^2 -(r^2+a^2) a \Phi}{a \Delta} + \Phi \right] K(k)\bigg\rbrace. 
%\end{eqnarray}

%\subsection{Selected orbit for timelike particle around KN black hole}

\subsubsection{\bf Selected Orbit}

\begin{figure}%[H]
\centering
\includegraphics[scale=0.6]{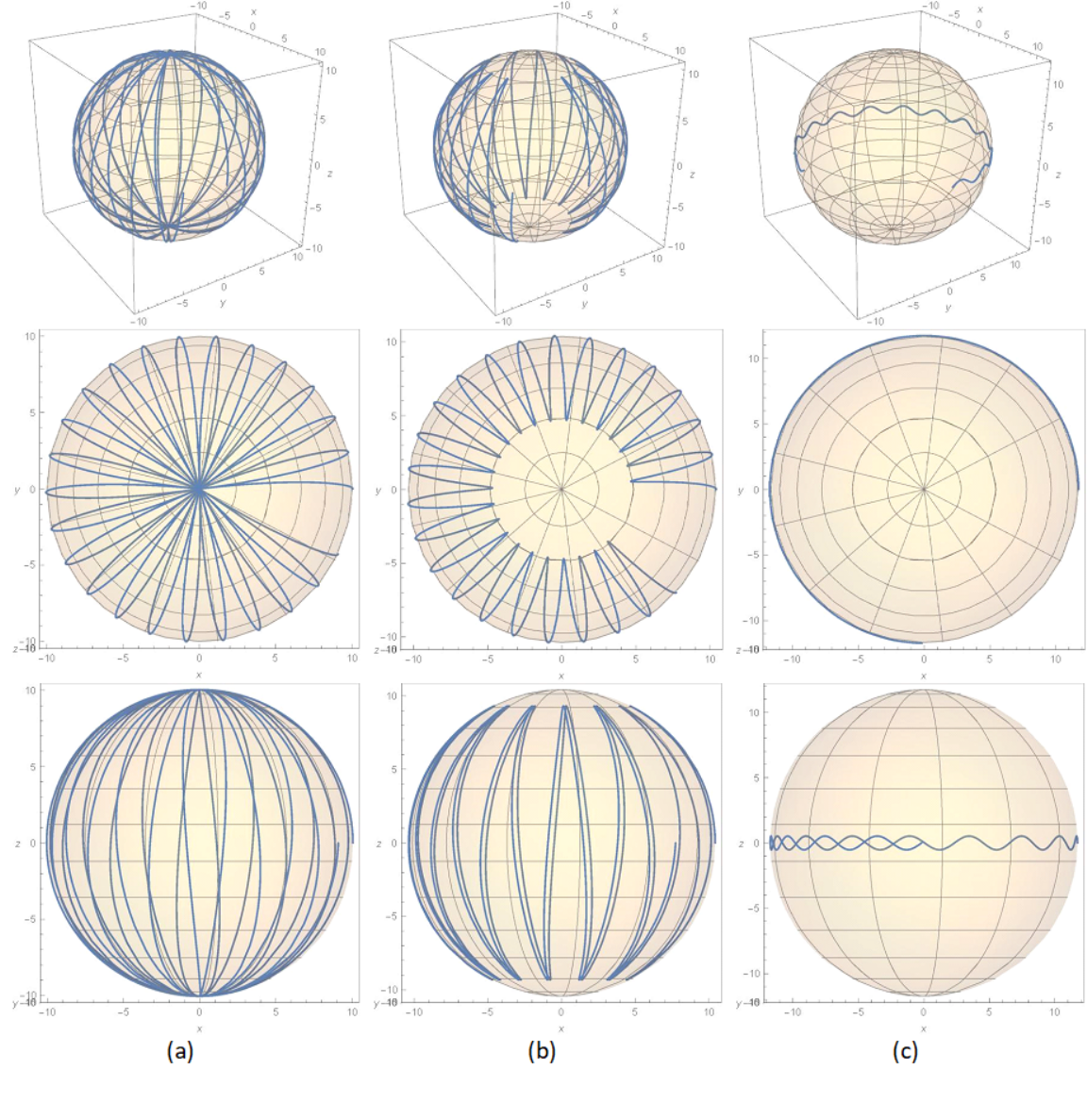}
\caption{Spherical timelike orbits of extremal KN with $\Phi = 0$ and $C=14.1 M$ in the $(x-y-z)$ as well as the projective $x-y$ and $x-z$ planes. The orbit begins from the equator and heads northwards. The direction of the black hole’s rotation is from west to east. Figures (a), (b), and (c) are cases with $Q=0$, $Q=0.45M$, and $Q=0.999M$, respectively.}
\label{fig:Phi0time}
\end{figure}

The analytical solutions described by Eqs.~\eqref{boundtimeKN1}-\eqref{boundtimeKN3} offer insight into the behavior of STOs. We will focus on plotting the case of extreme Kerr-Newman (KN) only, as the non-extreme scenario exhibits similar qualitative behavior.
%The analytical solutions~\eqref{boundtimeKN1}-\eqref{boundtimeKN3} can be plotted to show how the STO behaves. Here we shall plot the case of extreme KN only, as the non-extreme one has the same qualitative behavior. 
\begin{table}
\centering
\begin{tabular}{||c c c c c c||} 
 \hline\hline
\ \ Orbit\ \ &\ \ $Q/M$\ \ &\ \ $r_{SO}/M$\ \ &\ \ E\ \ &\ \ $u_0$\ \ &\ \ $\Delta\phi$\ \ \\ [0.5ex] 
 \hline
\ \ (a)\ &\ \ 0\ \ &\ \ 10\ \ &\ \ 0.9559\ \ &\ \ 1\ \ &\ \ 0.4869\ \ \\ 
\ \ (b)\ \ &\ \ 0.45\ \ &\ \ 10.39\ \ &\ \ 0.9570\ \ &\ \ 0.8930\ \ &\ \ 0.4624\ \ \\
\ \ (c)\ \ &\ \ 0.999\ \ &\ \ 11.74\ \ &\ \ 0.9606\ \ &\ \ 0.045\ \ &\ \ 0.3915\ \ \\
 \hline\hline
\end{tabular}
\caption{Parameters for STO around extremal KN with $\Phi=0$ and  $C= 14.1 M$.}
\label{tab5}
\end{table}
In Fig.~\ref{fig:Phi0time}, we present three examples of orbits corresponding to $\Phi=0$. Each plot depicts the orbit on an imaginary sphere with a fixed radius (its value is provided in Table~\ref{tab5}). The particle initiates its movement from the equator and progresses northward in each orbit, in alignment with the rotation direction of the black hole itself, which is from west to east.% In Fig.~\ref{fig:Phi0time}, we illustrate three examples of orbits for $\Phi=0$. In each plot, the orbit is plotted on an imaginary sphere with a fixed radius (its value is given in Table~\ref{tab5}). The particle begins to move from the equator and heads north in each orbit. The direction of rotation of the black hole itself is from west to east.
\begin{table}
\centering
\begin{tabular}{||c c c c c c||} 
 \hline\hline
\ \ Orbit\ \ &\ \ $Q/M$\ \ &\ \ $r_{SO}/M$\ \ &\ \ E\ \ &\ \ $u_0$\ \ &\ \ $\Delta\phi$\ \ \\ [0.5ex] 
 \hline
\ \ (a)\ \ &\ \ 0\ \ &\ \ 7\ \ &\ \ 0.9500\ \ &\ \ 0.9313\ \ &\ \ -46.4684\ \ \\ 
\ \ (b)\ \ &\ \ 0.45\ \ &\ \ 8.5361\ \ &\ \ 0.9526\ \ &\ \ 0.8317\ \ &\ \ -14.3344\ \ \\
\ \ (c)\ \ &\ \ 0.999\ \ &\ \ 11.35\ \ &\ \ 0.9595\ \ &\ \ 0.0416\ \ &\ \ -50.765\ \ \\
 \hline\hline
\end{tabular}
\caption{STO parameters around the extremal Kerr-Newman black hole with $\Phi=-1.350M$ and $C=12M$.}
\label{tab6}
\end{table}
As the charge increases, the orbits progressively concentrate more around the equator. When $Q= 0.999M$ (approaching the limit $Q\rightarrow M$) the orbit effectively becomes equatorial. 

\begin{figure}%[H]
\centering
\includegraphics[scale=0.6]{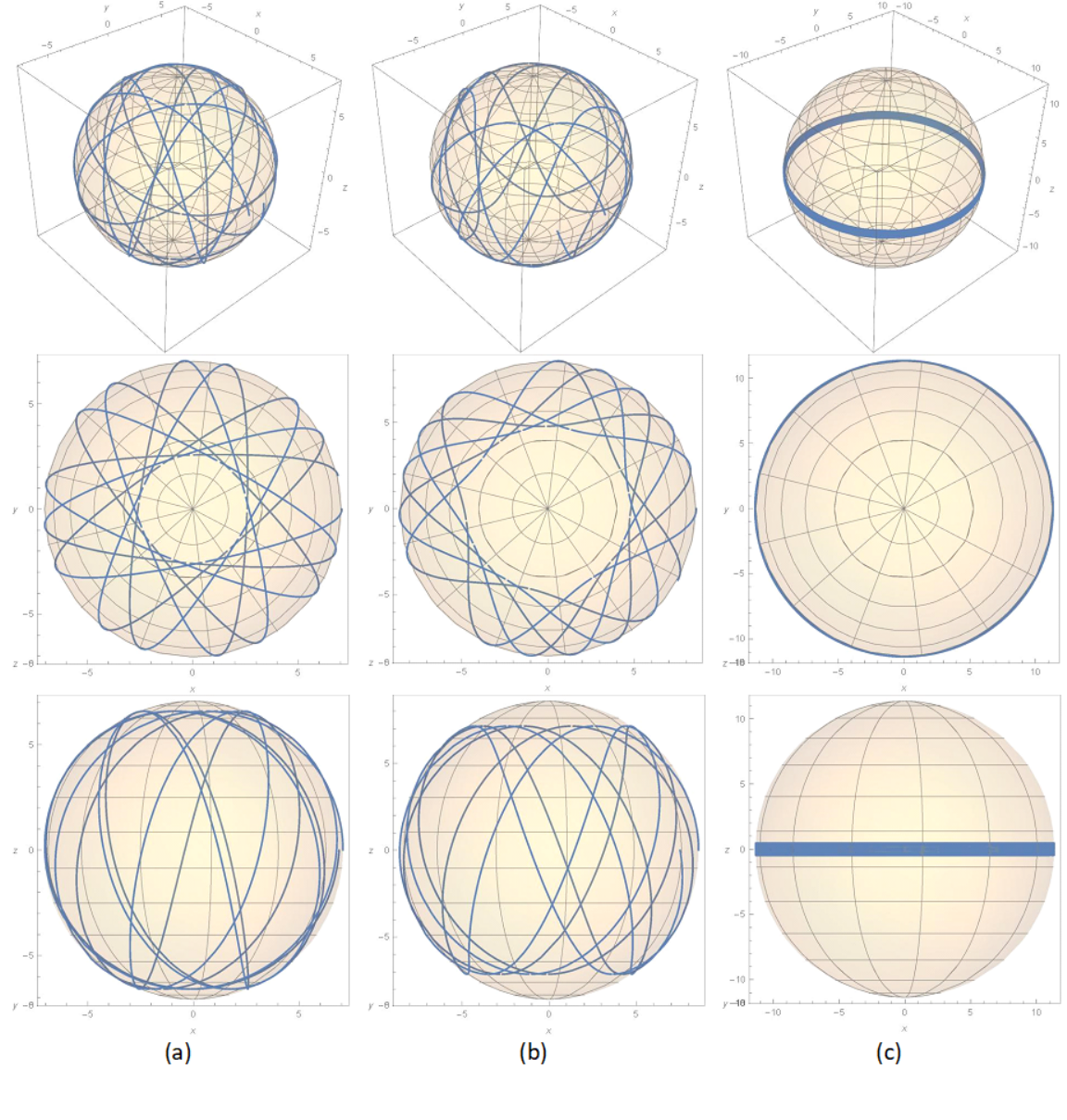}
\caption{Spherical orbit for extremal KN with $\Phi = -1350M$ and $C=12M$ in $(x-y-z)$ plane, the projective $x-y$ and $x-z$ plane. The orbit begins from the equator and heads northwards. The direction of the black hole’s rotation is from west to east. Label (a), (b), and (c) are cases when the charged value is $Q=0$, $Q=0.45M$, and $Q=0.999M$ respectively.}
\label{fig:Phi-136}
\end{figure}

In Fig.~\ref{fig:Phi-136} we depict STOs for $\Phi=-1.350M$, with the radii and other orbit parameters detailed in Table\ref{tab6}. These orbits are plotted for $Q=0$, $Q=0.45M$, and $Q=0.999M$, respectively. As the charge $Q$ increases, it not only alters the orbital direction of the test particle but also confines the orbit closer to the equatorial plane. However, unlike photon orbits, we observe that in the timelike case, there is no transition from retrograde to prograde motion with increasing charge.% also restricts the orbit to be equatorial. However, unlike the photon orbit, we found that the timelike case does not change from retrograde to prograde.

\subsubsection{\bf Innermost Stable Spherical Orbits}

\begin{figure}%[H]
\centering
\includegraphics[scale=0.6]{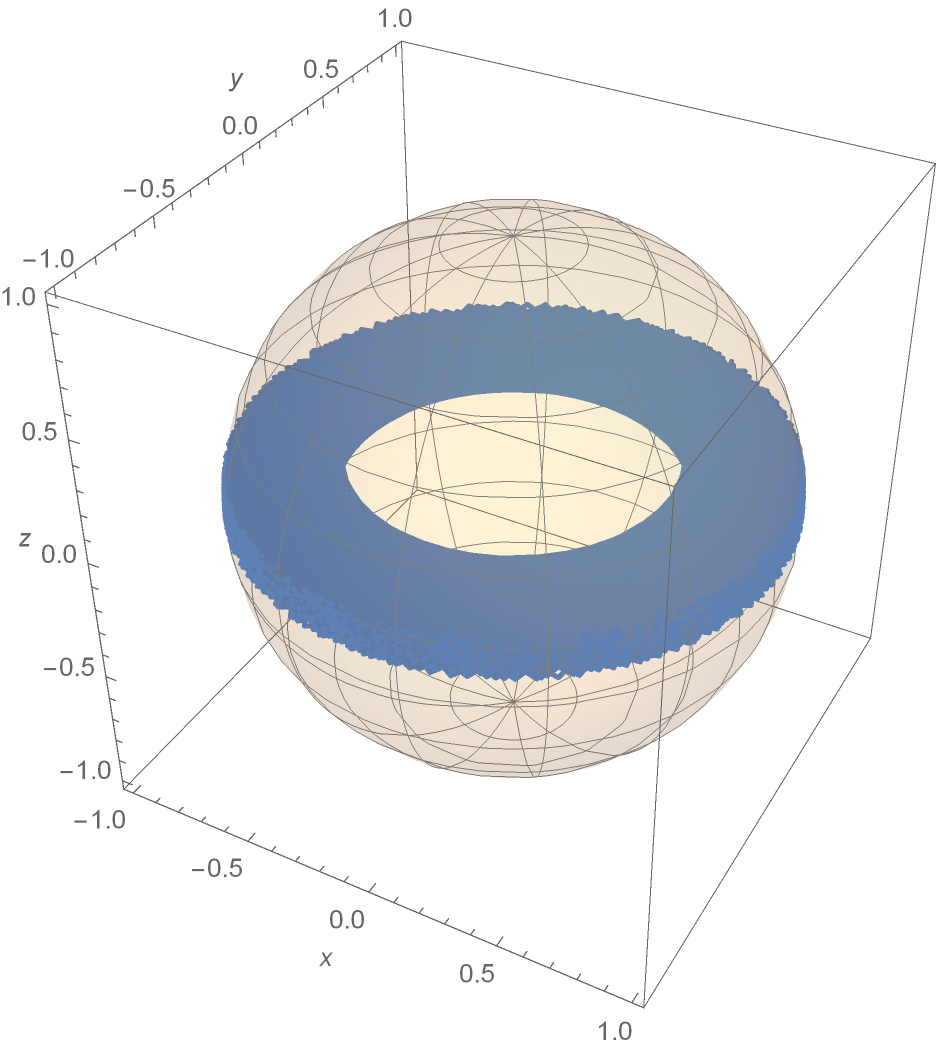}
\caption{ISSO in one-thirtieth second latitude oscillation when $Q=0.45M$ and $\Phi=0.192M$ for the extremal KN, having the same energy as in Table~\ref{tab6} (b). The corresponding radius is $r_{ISSO}=1.00085M$. The plot will fill the spherical coordinate as it reaches one latitudinal oscillation.}
\label{fig:pi16}
\end{figure}

From Eq.\eqref{KNgeod2}, it is evident that the function $R(r)$ in~\eqref{extraKN} serves as the effective radial potential, $V_{eff}(r)\equiv-R(r)$. Analogous to static cases, extrema typically exist for each specific choice of parameters, representing stable or unstable spherical orbits. At certain critical parameter values, these extrema merge. In the context of equatorial motion, this phenomenon is referred to as the "innermost stable circular orbit" (ISCO), while for spherical motion, it is termed the "{\it innermost stable spherical orbit}" (ISSO)~\cite{Druart:2023mba}. The ISSO is defined as:
\begin{equation} V_{eff}(r)\bigg|_{r=r_{ISSO}}=\frac{dV_{eff}(r)}{dr}\bigg|_{r=r_{ISSO}}=\frac{d^2V_{eff}(r)}{dr^2}\bigg|_{r=r_{ISSO}}=0.  
\end{equation}
This condition, supplemented with Eq.~\eqref{KNreq}, yields
\begin{eqnarray}
\label{KNisso1}
\frac{-\gamma +\delta +f+6 M r \sqrt{\sigma }-8 Q^2 \sqrt{\sigma }}{r T \left(M r-Q^2\right)}=0,\\
\label{KNisso2}
\frac{V \left(\gamma +\delta -f+6 M r \sqrt{\sigma }-8 Q^2 \sqrt{\sigma }\right)}{T}=0,
\end{eqnarray}
for the $(\pm)$ conditions,  respectively. Here,
\begin{eqnarray}
    \delta &\equiv& S \left(8 r_{ISSO}^2 \left(C \left(3 M^2+Q^2\right)-Q^4\right)-26 C M Q^2 r_{ISSO} + 8CQ^4 + U\right),\nonumber\\
    \gamma &\equiv& 2 a^4 C r_{ISSO}^2 \left(4Q^2 - 5Mr_{ISSO}\right) \left(Q^2-M r_{ISSO}\right),\nonumber\\
    V &=& a r_{ISSO} \left(a^2 \eta + r_{ISSO} \left(r_{ISSO} (r_{ISSO} - 2M) + Q^2\right)^2\right),\nonumber\\
    T &\equiv& 4 a^2 \left(Q^2-M r_{ISSO}\right)+\left(r_{ISSO} (r_{ISSO} - 3M) + 2Q^2\right)^2,\nonumber\\
    f &\equiv& \xi r_{ISSO}^4 \left(Q^2-M r_{ISSO}\right)  \left(r_{ISSO} (r_{ISSO}-3 M)+2 Q^2\right),\nonumber\\
    \sigma &\equiv& S^2 \left(a^2+r_{ISSO} (r_{ISSO}-2 M)+Q^2\right)^2 \zeta,\nonumber\\
    S &\equiv& a^2 r_{ISSO}^2 \left(Q^2-M r_{ISSO}\right),\nonumber\\
    U &\equiv& r_{ISSO}^4 \left(4C - 7M^2 + 4Q^2\right)+2 M r_{ISSO}^3 \left(7Q^2 - 9 C\right) - 3Mr_{ISSO}^5,\nonumber\\
    \xi &\equiv& (-9 M Q^2 r_{ISSO}+M r_{ISSO}^2 (6 M-r_{ISSO})+4 Q^4,\nonumber\\
    \zeta &\equiv& (a^2 C^2+r_{ISSO}^2 \left(3 C M r_{ISSO}-r_{ISSO}^2 \left(C+Q^2\right)-2 C Q^2+M r_{ISSO}^3\right),\nonumber\\
    \eta &\equiv& C (M-r_{ISSO})+r_{ISSO} \left(Q^2-M r_{ISSO}\right).\nonumber\\
\end{eqnarray}
\begin{figure}%[H]
\centering
\includegraphics[scale=0.6]{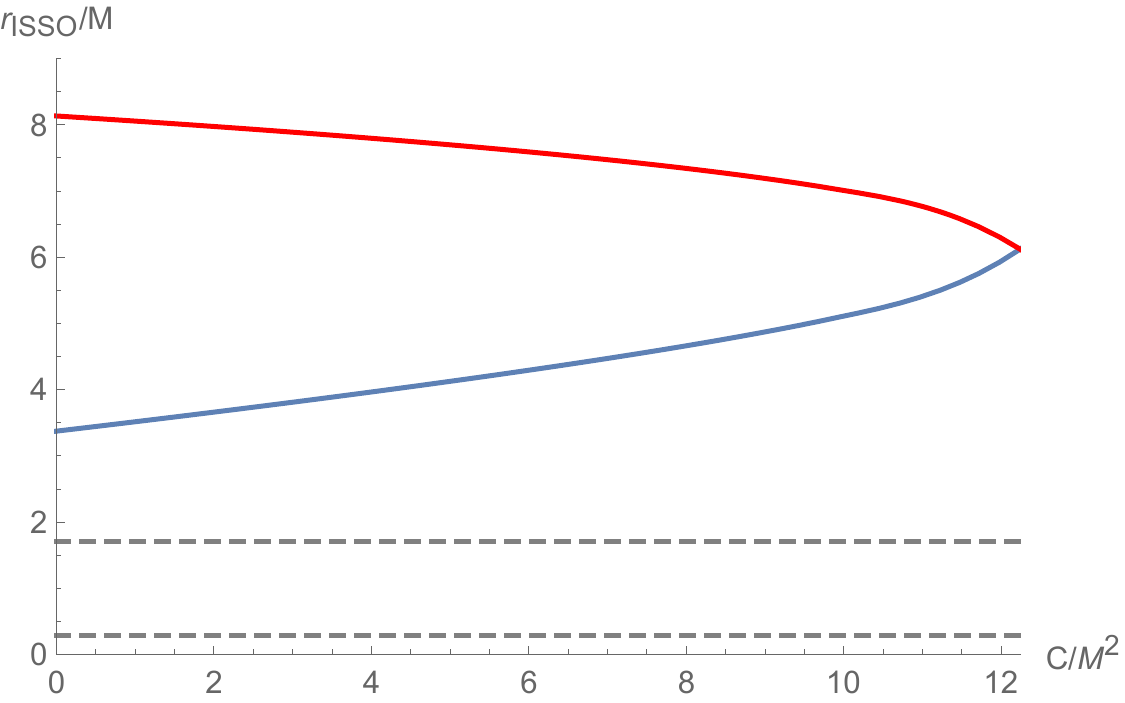}
\caption{Plot of $r_{isso}$ as a function of $C$ for $a=0.7M$ and $Q=0.1M$. The dashed lines are the corresponding horizons. The blue (red) lines are the negative (positive) roots of Eq.~\eqref{KNisso1} (Eq.~\eqref{KNisso2}), respectively.}
\label{fig:rissokn}
\end{figure}
We solve these equations to derive $r_{ISSO}$ as a function of Carter constant. The resulting plot is depicted in Fig.~\ref{fig:rissokn}, aligning with the findings of Wang, Lee, and Lin~\cite{Wang:2022ouq}. For each fixed value of $a$ and $Q$, two branches of observable $r_{ISSO}$($C$) exist outside the horizons. The upper branch originates at $r_{ISSO}/M\sim8.13$, while the lower branch begins at $r_{ISSO}/M\sim3.39$. These branches merge at a critical value $r_{ISSO}/M\sim6$, corresponding to $C_{crit}/M^2\sim12.23$, beyond which no $r_{ISSO}$ exists. The associated ISSO trajectory is illustrated in Fig.~\ref{fig:pi16}, with $r_{ISSO}=1.00085M$. While the trajectory may initially appear dense around the equator, it is important to note that the figure only depicts one-thirtieth of a complete latitudinal oscillation. A simulation of a full latitudinal oscillation reveals that the orbit fills the entire solid angle.

%. Instead, for the extremal KN the $V_{eff}$ eventually goes monotonically upward leaving one unique absolute minimum,
%\begin{equation} \frac{d^2V_{eff}}{dr^2}\bigg|_{r=r_*}<0.  
%\end{equation}
%We call this phenomenon the {\it innermost absolutely stable spherical orbits} (IASSO). Its corresponding spherical radius is located close to its event horizon $(r_*=1.00085M)$. This happens at $\Phi=1.92M$ with $E=0.9526$ (the same as in \ref{tab6} (b)).  The corresponding IASSO trajectory is shown in Fig.~\ref{fig:pi16}. The trajectory appears to look dense around the equator, but note that here we just show the one-thirtieth latitudinal oscillation. Simulation of one-full latitudinal oscillation shows that the orbit fills the whole solid angle.}

%\begin{figure}%[H]
%\centering
%\includegraphics[scale=0.7]{KH ISSO.pdf}
%\caption{Potential effective plot of $V_{eff}(r)=-R(r)$ shifts before and after merged (ISSO). Where dashed and thick plots are before at $\Phi=0$ and after merging at $\Phi=1.92M$ with other parameters from table \ref{tab6} (b), respectively, and the dotted line is the event horizon.}
%\label{fig:ISSOKH}
%\end{figure}

%\textcolor{red}{With the equation $\frac{d^2V_{eff}(r)}{dr^2}\bigg|_{r=r_{ISSO}}=0$ and substituting equation \eqref{KNreq}, we will get two solutions}

%\textcolor{red}{where the equation \eqref{KNisso1} and \eqref{KNisso2} are the solution from equation \eqref{KNreq} $(\pm)$ conditions respectively, and variables are}

%\textcolor{red}{We can solve the equation \eqref{KNisso1} and \eqref{KNisso2} to get $r_{ISSO}$ as a function of Carter constant in Fig \ref{fig:rissokn}.} 

%\newpage

\section{Spherical Orbits around Regular Black Hole}
\label{sec:regu}

In 1968, Bardeen introduced a regular metric that lacked a physical singularity, exhibiting regularity in both the metric and all invariants as $r\rightarrow0$~\cite{Bardeen}. However, Bardeen did not explicitly detail the method used to derive this solution in his original work. Later, Ayon-Beato and Garcia provided an explanation, demonstrating that such a solution can be sourced by a nonlinear electrodynamics (NLED) charge~\cite{Ayon-Beato:1998hmi, Ayon-Beato:2000mjt}. Since then, numerous proposals have emerged regarding nonsingular black holes coupled with NLED. The first to extend this regularization approach to the case of a rotating metric in Boyer-Lindquist coordinates was Ghosh~\cite{Ghosh:2014pba}, who derived a class of three-parameter axisymmetric metrics:
%In 1968 Bardeen presented a regular metric devoid of physical singularity, where the metric and all invariants are regular at $r\rightarrow0$~\cite{Bardeen}. In his original paper, Bardeen never showed how the solution was obtained. Ayon-Beato and Garcia later provided an explanation that such a solution can be sourced by NLED charge~\cite{Ayon-Beato:1998hmi, Ayon-Beato:2000mjt}. Ever since there have been many proposals on nonsingular black holes coupled with NLED. Balart and Vagenas~\cite{Balart:2014cga} and later Culetu~\cite{Culetu:2014lca} employ the distribution function inspired by the shape of the probability density function to regularize the static metric. The first to generalize it to the case of rotating metric in Boyer-Lindquist coordinate is Ghosh \cite{Ghosh:2014pba} who obtained a class of three-parameter axisymmetric metric:
\begin{eqnarray}
\label{regrot}
ds^2=-\left(1-\frac{2m(r)r}{\Sigma}\right){dt}^2-\frac{4am(r)r}{\Sigma}\sin^2\theta dt\ d\phi
+\Sigma\left(\frac{{dr}^2}{\Delta}+{d\theta}^2\right)+\frac{\mathcal{A}}{\Sigma}{\sin^2\theta d\phi}^2,\nonumber\\
\end{eqnarray}
where
\begin{eqnarray}
\label{regrotdetail}
m(r)&\equiv&Me^{-k/r},\nonumber\\
\Sigma&\equiv& r^2+a^2\cos^2\theta,\nonumber\\ %\label{regrotdetail1}\\
\Delta&\equiv&r^2+a^2-2m(r)r,\nonumber\\% \label{regrotdetail2}\\
\mathcal{A}&\equiv&\left(r^2+a^2\right)^2-{\Delta a}^2\sin^2\theta.%\label{regrotdetail3}
\end{eqnarray}
The metric depends on the mass $M$, the angular momentum $a$, and a new parameter $k$ which controls its regularity. The range of $a$ is between $0<a^2 \leq M^2$, where the upper bound is related to the extreme condition or even "naked" (depending on the value of k) and the lower bound is Schwarzchild black hole (for $k=0$). The event horizons $r_{\pm}$ satisfy
\begin{eqnarray}
\label{deltaregrot}
\begin{aligned}
r_{\pm}^2+a^2-2m(r_{\pm})=0.
\end{aligned}
\end{eqnarray}
Here the radii $r_{\pm}$ can only be obtained numerically.

Ghosh obtained the solution~\eqref{regrot} by transforming the non-singular static black holes by the shape of the probability density function~\cite{Balart:2014cga, Culetu:2014lca} to the rotating solutions via the Newman-Janis algorithm~\cite{Newman:1965tw}. At the core ($r\rightarrow0$) all the invariants of this metric are regular everywhere. Specifically, the {\it Kretschmann} scalar ($K$) approaches zero there, $K(r\rightarrow0)\rightarrow0$.  For $k=0$ the metric reduces to Kerr (when $a\neq0$) or Schwarzschild (when $a=0$). In the non-rotating limit ($a\rightarrow0$) the metric reduces to the regular spherically-symmetric black hole~\cite{Balart:2014cga, Culetu:2014lca}
\begin{equation}
\label{regbalart}
ds^2=-\left(1-\frac{2M e^{-k/r}}{r}\right)dt^2+\left(1-\frac{2M e^{-k/r}}{r}\right)^{-1}dr^2+d\Omega_2^2.
\end{equation}
Interestingly, the spacetime structure at the core is (as long as $r\gg k$) Minkowski/RS-like,
\begin{equation}
 \left(1-\frac{2M e^{-k/r}}{r}\right)\approx1-\frac{2M}{r}+\frac{2Mk}{r^2}+\mathcal{O}\left(r^{-3}\right),   
\end{equation}
rather than de Sitter-like, as in the Bardeen case~\cite{Bardeen, Ayon-Beato:2000mjt, Ramadhan:2023ogm}.

% We are interested in investigating the orbit outside the outer horizon, $r_+<r<\infty$.}

\subsection{Spherical Photon Orbits}
\label{sec:SPOR}

\subsubsection{\bf Equation of Motion}

The four geodesic equations of photons in regular rotating space-time are
%\begin{subequations}
%\label{geodesicregrot}
\begin{eqnarray}
%\label{geodesicregrot}
\Sigma\Delta\dot{t}&=&\mathcal{A}E-2a m(r)rL_{z},\label{geodesicregrot1}\\
\Sigma\Delta\dot{\phi}&=&2am(r)rE+\left(\Sigma-2m(r)r\right)\frac{L_{z}}{\sin^2\theta},\label{geodesicregrot2}\\
\Sigma^2\dot{\theta}&=&\mathcal{C}-\left[\frac{L_{z}}{\sin^2\theta}-E^2a^2\right]\cos^2\theta,\label{geodesicregrot3}\\
\Sigma^2\dot{r}&=&E^2r^4+\left(E^2a^2-L_{z}^2-\mathcal{C}\right)r^2+2m(r)r\left[\left(aE-L_{z}\right)^2+\mathcal{C}\right]-a^2 \mathcal{C}.\label{geodesicregrot4}
\end{eqnarray}
%\end{subequations}
%The dot sign is the derivative with respect of affine parameter $\tau$ along the geodesic. 
The notation of $E$, $L_z$, and $\mathcal{C}$ are the same three constants of motion-defined previously.%}% $E$ and $L_z$ are energy of photon and angular momentum about $\phi$-axis, and $\mathcal{C}$ is Carter's constant.

The geodesic equation of $\theta$ has a similar form with Kerr.
Setting $u=\cos\theta$, Eq.~\eqref{geodesicregrot3} can be rewritten as
\begin{equation}
\label{udotsquared}
\left(\frac{\Sigma}{E}\right)^2\dot{u}^2=C-\left(C+\Phi^2-a^2\right)u^2-a^2u^4,
\end{equation}
where the new parameters are
\begin{equation}
\Phi\equiv\frac{L_{z}}{E}\:,\:C\equiv\frac{\mathcal{C}}{E}.
\end{equation}
The physical requirement is also similar to Kerr's black hole. When $C>0$, the positive root of~\eqref{udotsquared} is
\begin{equation}
    u_0^2 = \frac{(a^2 - C - \Phi^2) + \sqrt{(a^2 - C - \Phi^2)^2 + 4 a^2 C}}{2a^2}.
\end{equation}
For $C<0$,
\begin{equation}
\label{crestriction}
    a^2 - C - \Phi^2 > 0.
\end{equation}
%This condition is also restrictive and would rule out a photon orbit case  \cite{teo}.}

\subsubsection{\bf Requirement for the Existence of SPO}

Solving the two conditions resulting from Eq.~\eqref{socond} simultaneously we get
\begin{eqnarray}
(i) ~~\Phi & = &\frac{r_{SO}^2 + a^2}{a}, ~~ C ~~= -\frac{r_{SO}^4}{a^2}; \label{phic1} \\
(ii) ~~\Phi & = & \frac{a^2 k M+a^2 r_{SO}^2 e^{k/r_{SO}}+a^2 M r_{SO}+k M r_{SO}^2+r_{SO}^4 e^{k/r_{SO}}-3 M r_{SO}^3}{a \left(k M-r_{SO}^2e^{k/r_{SO}}+M r_{SO}\right)}, \nonumber\\ %\nonumber
C & = & -\frac{r_{SO}^4 \bigg\lbrace{4 a^2 M e^{k/r_{SO}} (k-r_{SO})+\left[r_{SO} \left(r_{SO} e^{k/r_{SO}}-3 M\right)+k M\right]^2\bigg\rbrace}}{a^2 \left[r_{SO} \left(M-r_{SO} e^{k/r_{SO}}\right)+k M\right]^2}.\label{phic2}
\end{eqnarray}
As in the KN case, both possibilities satisfy condition~\eqref{socond2} for all $r_+<r_{SO}$. Thus, all SPOs here are also unstable. Solution~\eqref{phic1} can be ruled out, and so the acceptable one is~\eqref{phic2} with $C>0$ to satisfy condition~\eqref{crestriction}.
\begin{figure}%[h]
\centering
\begin{tabular}{cc}	\includegraphics[width=0.4\linewidth]{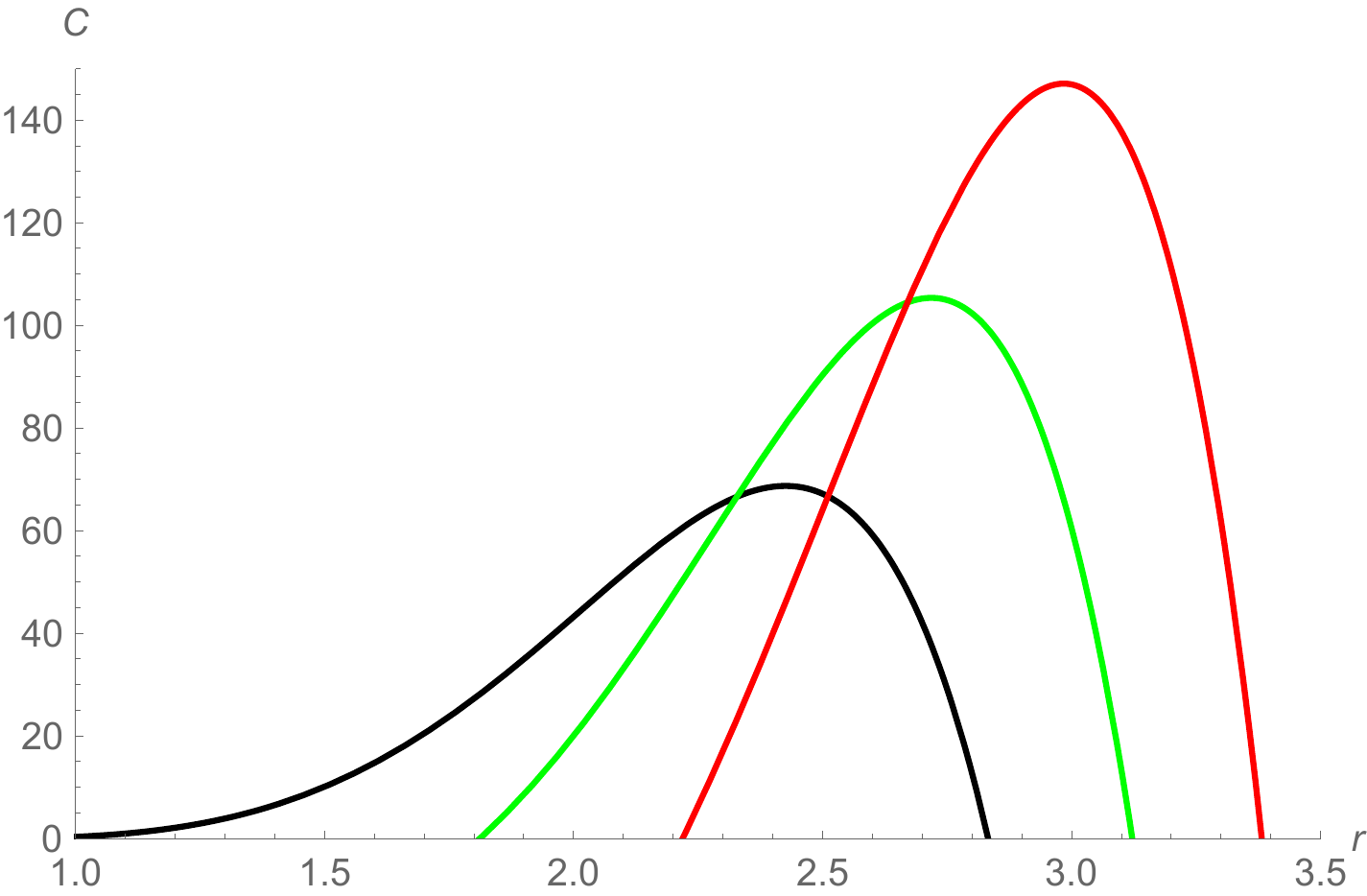} &
\includegraphics[width=0.4\linewidth]{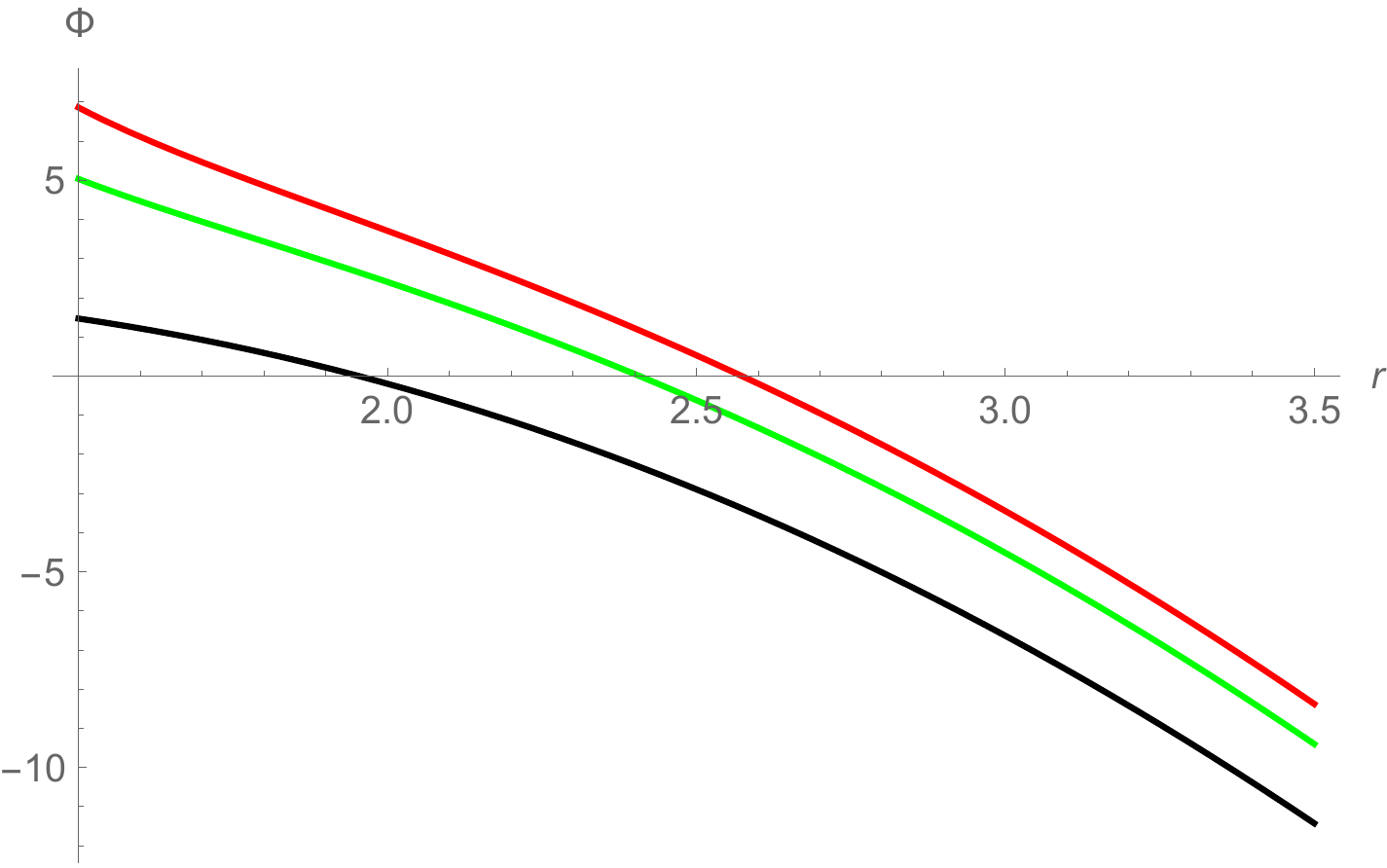}\\
\end{tabular}
\caption{Typical plots of $C$ and $\Phi$ as functions of $r_{SO}$ for several values of $k$: $k=0.1M$ (red), $k=0.3M$ (green), and $k=0.5M$ (black).}
\label{fig:regprop}
\end{figure}
The behavior of $C(r_{SO})$ and $\Phi(r_{SO})$ for Ghosh BH are depicted in Fig.~\ref{fig:regprop}. They look similar to the KN case, Fig.~\ref{fig:Shift}, but due to the nature of $m(r)$ the plots are not linear but bent around a small value of $r_{SO}$.

%The first $(i)$ solution is similar to the Kerr case, so it can immediately be ruled out as an unphysical case. These parameters must satisfy the constraint condition on equation \eqref{crestriction}. Since it is not satisfied in this case, the solution is ruled out. In the second $(ii)$ solution, $C$ may take positive or negative signs. However, condition \eqref{crestriction} is also unsatisfied when $C<0$. Thus, the only physical solution is when $C>0$ \cite{teo}.

\subsubsection{\bf Analytical Solutions}

Once again we employ the Mino formalism~\eqref{mino} to solve Eqs.~\eqref{geodesicregrot2}-\eqref{geodesicregrot4}. %The geodesic equation solved by using Mino's parameter $\lambda$ along the geodesic has the form $\frac{d\tau}{d\lambda}=\Sigma$. Thus, we can write
The geodesic equations for $\phi$ and $t$ can be written
\begin{eqnarray}
    \frac{d \phi}{dw} &=& \frac{1}{2Y(w)} \left[\frac{2am(r)r-a^2\Phi}{\Delta}+\frac{\Phi}{\left(1-w\right)}\right], \\
    \frac{d t}{dw} &=&  \frac{1}{2Y(w)} \left[ \frac{(r^2+a^2)^2 -2a \Phi m(r)r }{\Delta} -a^2(1-w), \right]
\end{eqnarray}
where $Y(w)$ is the same as in Eq.~\eqref{yw}. Assuming the initial condition $\phi=t=0$ when $w=0$, the analytic solution of $\phi$ and $t$ can be written as
\begin{eqnarray}
    \phi(\lambda) &=& \frac{\Phi}{a(1-w_2)\sqrt{w_1-w_2}} [ F(\Psi,\kappa) - w_2 \Pi(\Psi,\kappa^2(1-w_2),\kappa) ] \\ \nonumber
    &&+ \frac{a}{\Delta}\left(2m(r)r-a\Phi\right)\lambda, \\ 
    t(\lambda) &=& - \frac{a}{\sqrt{w_1-w_2}} [ (1-w_2) F(\Psi,\kappa) + w_2 \Pi(\Psi,\kappa^2,\kappa) ] \\ \nonumber
    &&+ \frac{1}{\Delta}\left[ (r^2+a^2)^2 - 2a\Phi m(r)r\right] \lambda, 
\end{eqnarray}
where $\Psi$ and $\kappa$ are given by Eq.~\eqref{psik}. The solution for $\theta$ is the same as in Eq.~\eqref{thetasol}.%where $\Pi(\Psi,w_1,\kappa)$ is the incomplete elliptic integral function of the third kind. The above two solutions are valid for all values of the $\lambda$ parameter. When $e^{-\frac{k}{r}}=1$, the solution will be reduced to Teo's solution.}

The changes in $\phi$ and $t$ for one complete latitudinal oscillation are
\begin{eqnarray} 
    \Delta \phi &=& \frac{4}{\sqrt{w_1-w_2}} \left[ \frac{\Phi}{a(1-w_1)} \Pi(-\frac{w_1}{1 - w_1},\kappa) +  \frac{2m(r)r - a \Phi }{\Delta}  K(\kappa) \right], \label{deltaphireg} \\ 
    \Delta t &=&  \frac{4}{\sqrt{w_1-w_2}} [ -a((1-w_2) K(\kappa) + w_2 \Pi(\kappa^2,\kappa))\nonumber\\
    &&+  \frac{(r^2+a^2)^2 -2a\Phi m(r)r}{a \Delta}  K(\kappa)]. 
\end{eqnarray}
The equation $\Delta \phi$ determines the orbit direction. Positive $\Delta\phi$ means prograde orbit, while negative one means retrograde orbit.

\subsubsection{\bf Selected Orbits}

Using the analytical solutions $(\theta,\phi,t)$ we will show typical orbits for extreme, and naked conditions of regular rotating black holes. Unlike the KN case, however, here we have an additional parameter to consider, $k$, which appears non-linearly in the Eq.~\eqref{deltaregrot},
\begin{equation}
\label{deltarescaled}
\left(\frac{r_{\pm}} {M}\right)^2+\left(\frac{a}{M}\right)^2-\frac{2r_{\pm}}{M}e^{-Mk/r_{\pm}}=0.
\end{equation}
Consequently, the extremality can be achieved by either adjusting $a/M$ or $Mk$. %Photon begins to move from the equator and head north in each orbit, while the direction of the black hole's rotation is from west to east. 
\\
\\

{\it Extremal Condition}
\begin{table}%[H]
\centering
\begin{tabular}{||c c c c c c c||} 
 \hline\hline
\ \ Orbit\ \ &\ \ $k$\ \ &\ \ $a/M$\ \ &\ \ $r_{SO}/M$\ \ &\ \ $C/M^2$\ \ &\ \ $u_0$\ \ &\ \ $\Delta\phi$\ \ \\ [0.5ex] 
 \hline
\ \ (a)\ \ &\ \ 0\ \ &\ \ 1\ \ &\ \ 2.414\ \ &\ \ 22.314\ \ &\ \ 1\ \ &\ \ 3.1761\ \ \\ 
\ \ (b)\ \ &\ \ 0.1\ \ &\ \ 0.899\ \ &\ \ 2.339\ \ &\ \ 21.124\ \ &\ \ 0.809\ \ &\ \ 3.7483\ \ \\
\ \ (c)\ \ &\ \ 0.5\ \ &\ \ 0.472\ \ &\ \ 1.953\ \ &\ \ 15.861\ \ &\ \ 0.222\ \ &\ \ 9.931\ \ \\
 \hline\hline
\end{tabular}
\caption{SPO parameters around an extremal Ghosh black hole with $\Phi=0$.}
\label{nullex1}
\end{table}
\begin{table}%[H]
\centering
\begin{tabular}{||c c c c c c c||} 
 \hline\hline
\ \ Orbit\ \ &\ \ $k$\ \ &\ \ $a/M$\ \ &\ \ $r_{SO}/M$\ \ &\ \ $C/M^2$\ \ &\ \ $u_0$\ \ &\ \ $\Delta\phi$\ \ \\ [1ex] 
 \hline
\ \ (a)\ \ &\ \ 0\ \ &\ \ 1\ \ &\ \ 2\ \ &\ \ 16\ \ &\ \ 0.971\ \ &\ \ 10.843\ \ \\ 
\ \ (b)\ \ &\ \ 0.1\ \ &\ \ 0.899\ \ &\ \ 1.946\ \ &\ \ 15.182\ \ &\ \ 0.785\ \ &\ \ 8.417\ \ \\
\ \ (c)\ \ &\ \ 0.5\ \ &\ \ 0.472\ \ &\ \ 1.675\ \ &\ \ 11.756\ \ &\ \ 0.214\ \ &\ \ 21.825\ \ \\ [1ex] 
 \hline\hline
\end{tabular}
\caption{SPO parameters around an extremal regular rotating black hole with $\Phi=M$.}
\label{nullex2}
\end{table}
\begin{figure}%[H]
\centering
\includegraphics[scale=0.45]{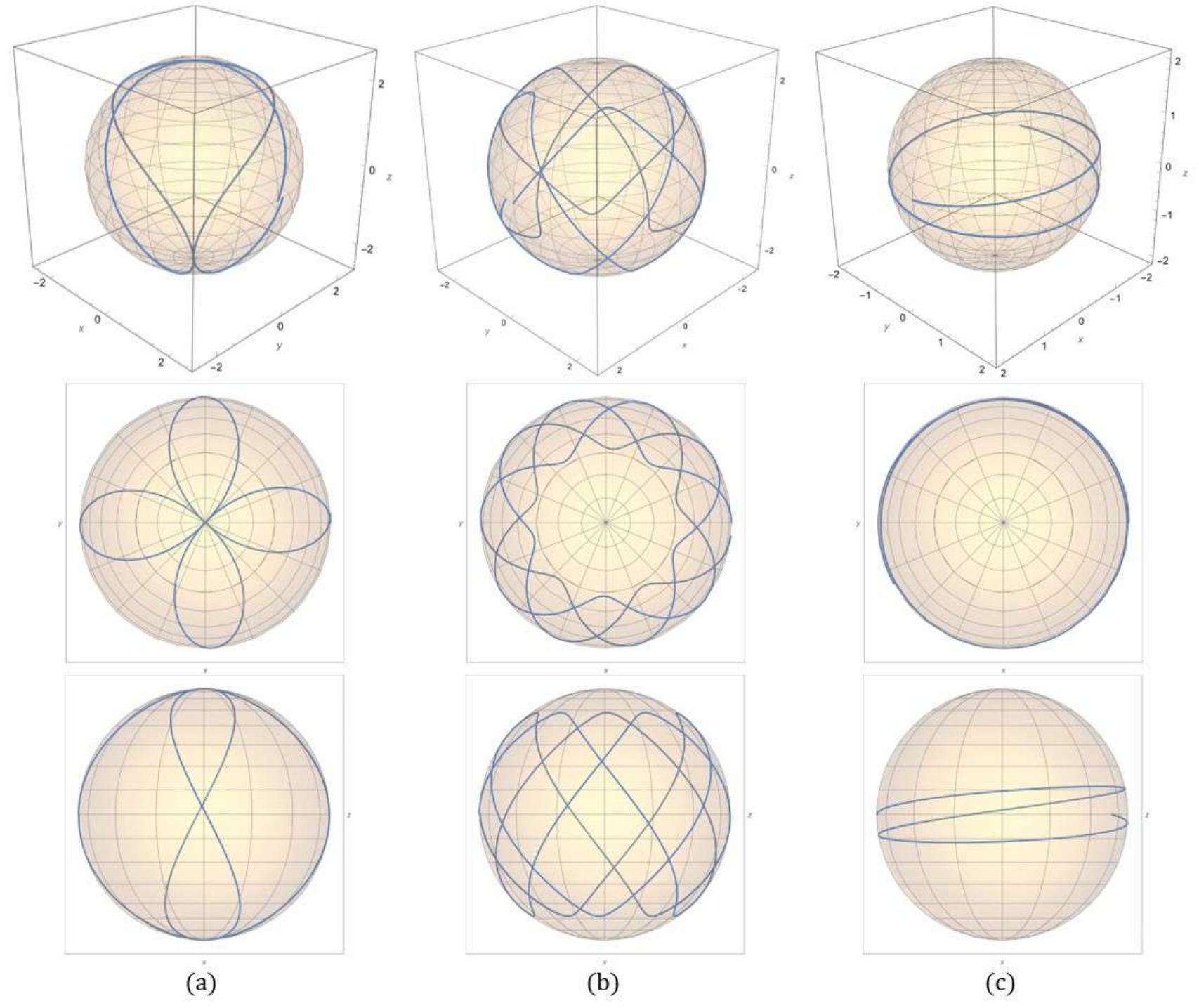}
\caption{Spherical photon orbits around extremal regular black hole with $\Phi = 0$ in the $(x-y-z)$ as well as the projective $x-y$ and $x-z$ planes. The direction of the black hole’s rotation is from west to east. Figures (a), (b), and (c) are cases with $k=0$, $k=0.1$, and $k=0.5$, respectively.}
\label{fig:ex1}
\end{figure}

\begin{figure}%[H]
\centering
\includegraphics[scale=0.45]{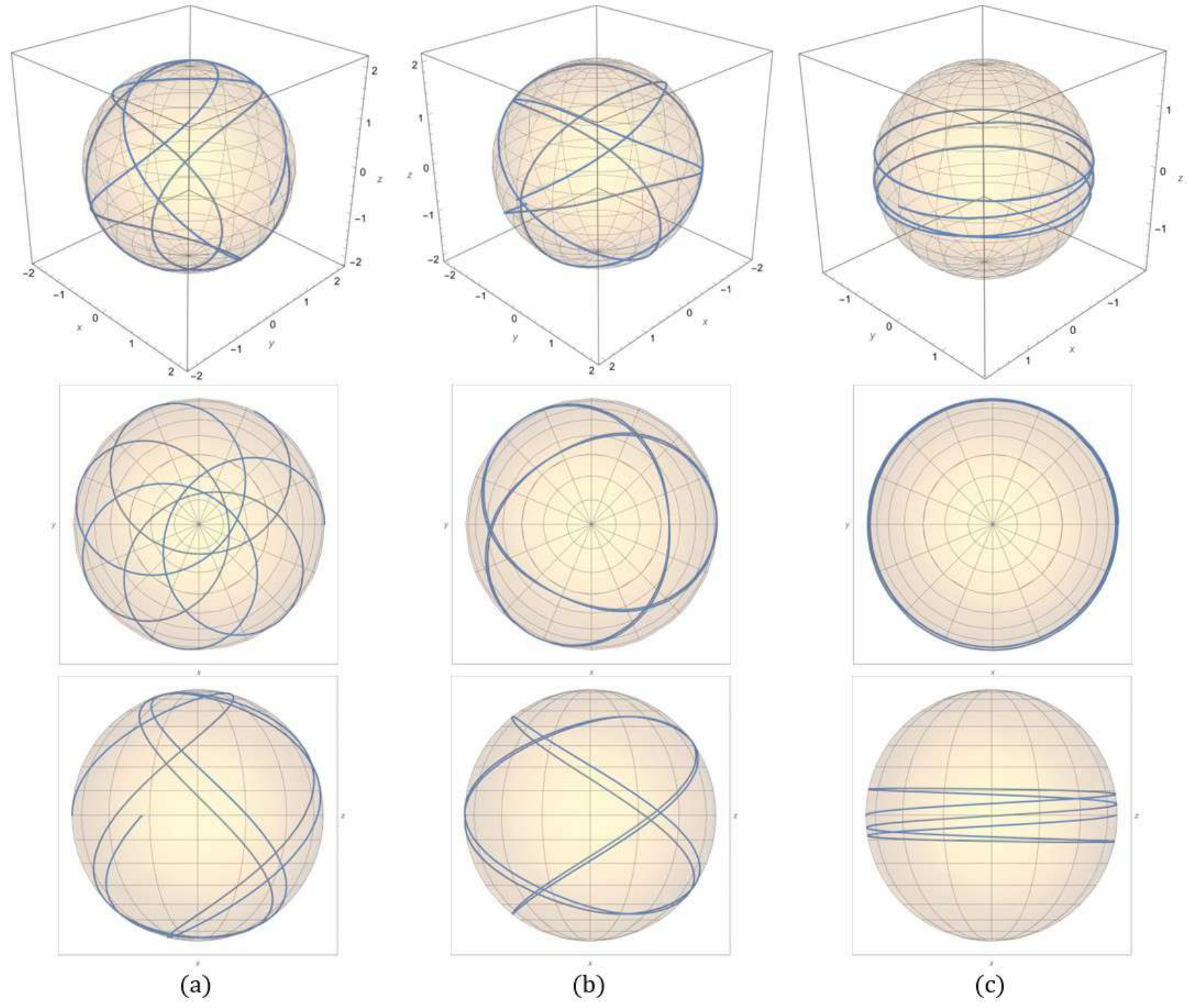}
\caption{Spherical photon orbits around extremal Ghosh black hole with $\Phi = M$ in the $(x-y-z)$ as well as the projective $x-y$ and $x-z$ planes. The direction of the black hole’s rotation is from west to east. Figures (a), (b), and (c) are cases with $k=0$, $k=0.1$, and $k=0.5$, respectively.}
\label{fig:ex2}
\end{figure}

The extremal condition is not compatible with $Mk>>1$, similar to the non-extremal case. There exists no parameter $a/M$ that satisfies:
\begin{enumerate}
\item $Mk>>1$, 
\item $r_{SO}>r_+\neq r_-$, and
\item Eq~\eqref{phic2},
\end{enumerate}
The extremal condition turns out to also be incompatible with $Mk>>1$ regime due to the same reasons as observed in the non-extremal scenario. In this study, we investigate the orbital behavior under the extremal case with $Mk=0,\ 0.1,$ and $0.5$. These cases are illustrated in Figs.~\ref{fig:ex1} (for $\Phi=0$) and~\ref{fig:ex2} (for $\Phi=M$), respectively, with the corresponding SPO parameters detailed in Tables~\ref{nullex1} and~\ref{nullex2}. For $k=0$, both SPOs ($\Phi=0,\ M$) reduce to the static one previously obtained by Teo in~\cite{teo}. %It can be seen in general that the shows orbit latitudinal oscillations. This oscillation has a property that varies due to changes in the values being varied, namely $k$ and $\Phi$. Changes in the value of $k$ cause the value of $a$ that must be changed to be minimized. The smaller a causes the amplitude of the latitudinal oscillations to get smaller, even approaching the equatorial orbit. Keep in mind again that at $a = 0$, the black hole will again be static and the orbit will be confined to the equatorial plane. 
For each value of fixed $\Phi$ it is evident that as $Mk$ increases, the orbits become increasingly confined to the equatorial plane. Moreover, with higher values of $\Phi$, the amplitude of the latitudinal oscillations diminishes, leading to longer oscillation periods. %, in other words, in one rotation there are less frequent oscillations, and the last one causes the revolution to be faster. This is because the value of $\Phi$ is the angular momentum of the photon so if it gets smaller, the revolution will slow down.}

\begin{figure}%[H]
\centering
\includegraphics[scale=0.36]{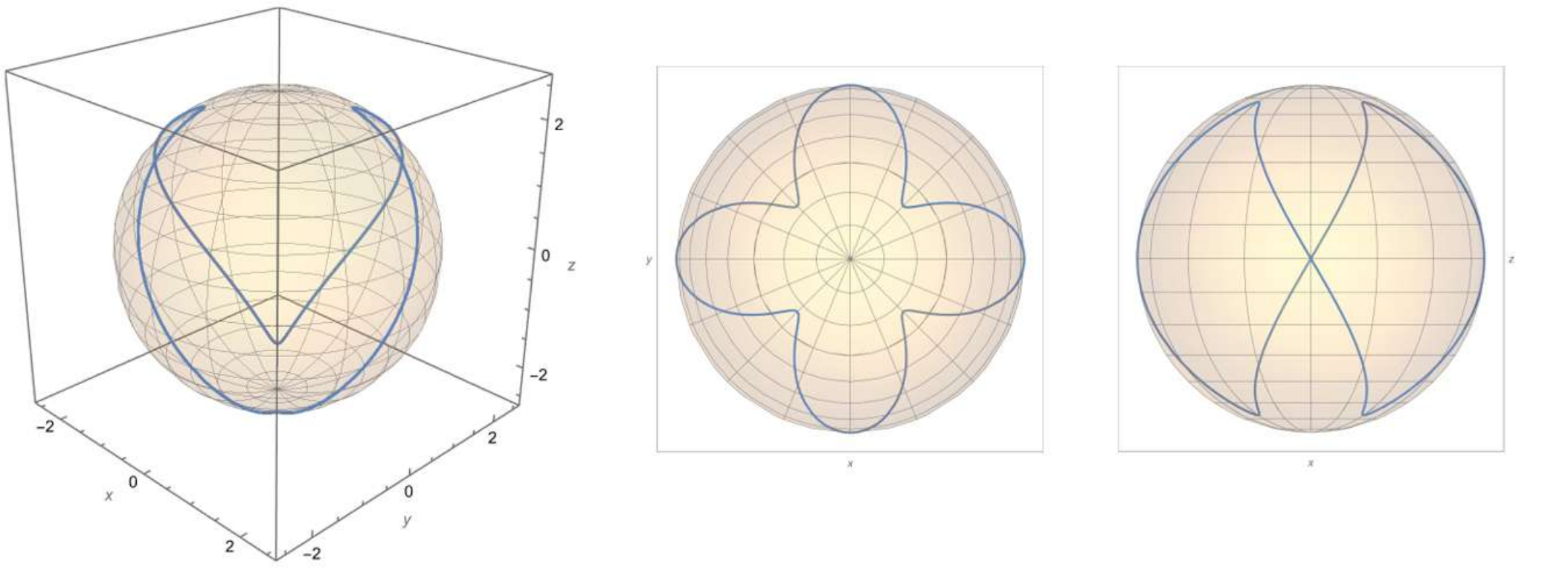}
\caption{Photon boomerang around extremal Ghosh black hole with $\Phi = -0.07$ in the $(x-y-z)$ as well as the projective $x-y$ and $x-z$ planes. The direction of black hole’s rotation is from west to east.}
\label{fig:boom1}
\end{figure}

In the extremal Ghosh black hole case, we also attempted to identify the \textit{photon boomerang}, characterized by the condition where Eq.~\eqref{deltaphireg} equals $\pi$. Through our numerical investigation, we found only an \textit{approximate} photon boomerang with $k=0.05$. We label it as ``approximate" because the value of $\Phi=0.07$ is non-zero. This is detailed in Table~\ref{nullexboom}, and the corresponding trajectory is depicted in Fig.~\ref{fig:boom1}.
%\begin{table}%[H]
%\centering
%\begin{tabular}{|| c c c c c c c||} 
% \hline\hline
%\ \ Orbit\ \ &\ \ $k$\ \ &\ \ $a/M$\ \ &\ \ $r_{SO}/M$\ \ &\ \ $C/M^2$\ \ &\ \ $u_0$\ \ &\ \ $\Delta\phi$\ \ \\ [0.5ex] 
% \hline
%\ \ (a)\ \ &\ \ 0.05\ \ &\ \ 0.94998\ \ &\ \ 2.402\ \ &\ \ 22.0487\ \ &\ \ 0.902\ \ &\ \ 3.1415\ \ \\ 
% \hline\hline
%\end{tabular}
%\caption{\textcolor{red}{Photon boomerang parameters around an extremal regular rotating black hole with $\Phi=-0.07$.}}
%\label{nullexboom}
%\end{table}
\begin{table}%[H]
\centering
\begin{tabular}{|| c c c c c c||} 
 \hline\hline
\ \ $k$\ \ &\ \ $a/M$\ \ &\ \ $r_{SO}/M$\ \ &\ \ $C/M^2$\ \ &\ \ $u_0$\ \ &\ \ $\Delta\phi$\ \ \\ [0.5ex] 
 \hline
\ \ 0.05\ \ &\ \ 0.94998\ \ &\ \ 2.402\ \ &\ \ 22.0487\ \ &\ \ 0.902\ \ &\ \ 3.1415\ \ \\ 
 \hline\hline
\end{tabular}
\caption{(Approximate) photon boomerang parameters around an extremal regular rotating black hole with $\Phi=-0.07$.}
\label{nullexboom}
\end{table}
%\clearpage

{\it No-horizon Spacetime}

Certain combinations of ${a/M,\ Mk}$ in Eq.\eqref{deltarescaled} can give rise to complex solutions for $r_{\pm}$, which, in a conventional black hole, is termed {\it naked singularity}. However, within the framework of a regular black hole, the concept of singularity vanishes. This scenario is often referred to as "{\it the no-horizon spacetime}" in the literature. While discussing trajectories around a naked singularity is nonsensical, the notion is perfectly acceptable in the case of a regular no-horizon spacetime.

%Certain combinations of $\{a/M,\ Mk\}$ in Eq.~\eqref{deltarescaled} can produce complex $r_{\pm}$. In an ordinary black hole, this is known as the naked singularity, which is (fortunately) unstable and unphysical by cosmic censorship theorem~\cite{Penrose:1964wq}. In a regular black hole, however, there is no singularity. This is commonly called ``{\it the no-horizon spacetime}" in the literature. While it is nonsensical to talk about trajectories around the naked singularity, it is perfectly legitimate in the case of regular no-horizon. 

The notion of a regular solution devoid of singularity is not a new concept in GR. Mazur and Mottola~\cite{Mazur:2004fk, Mazur:2001fv} discussed an alternative final state of non-singular gravitational collapse by invoking the Bose-Einstein condensate formalism in the interior. The consequence is a horizonless, regular core reminiscent of de Sitter space, which they coined as ``{\it gravastar}". More recently, there has been discourse surrounding the correlation between regular black holes and horizonless ultracompact stars~\cite{Carballo-Rubio:2019fnb, Carballo-Rubio:2022nuj, Cadoni:2022chn}. It is pertinent to distinguish that the Ghosh no-horizon solution differs conceptually from the gravastar, as the former features a Minkowski core while the latter exhibits a de Sitter-like core.%It is worth noting that the Ghosh no-horizon solution is conceptually different from the gravastar since the former has Minkowski core while the latter has de Sitter-like core.
\begin{table}%[H]
\centering
\begin{tabular}{||c c c c c c c c||} 
 \hline\hline
\ \ Orbit\ \ &\ \ $Mk$\ \ &\ \ $\Phi/M$\ \ &\ \ $a/M$\ \ &\ \ $r_{SO}/M$\ \ &\ \ $C/M^2$\ \ &\ \ $u_0$\ \ &\ \ $\Delta\phi$\ \ \\ [1ex] 
 \hline
\ \ (a)\ \ &\ \ 0.1\ \ &\ \ 0\ \ &\ \ 1\ \ &\ \ 1.242\ \ &\ \ 24.784\ \ &\ \ 1\ \ &\ \ 11.426\ \ \\ 
\ \ (b)\ \ &\ \ 0.1\ \ &\ \ -0.55\ \ &\ \ 1\ \ &\ \ 1.178\ \ &\ \ 36.224\ \ &\ \ 0.995\ \ &\ \ 6.291\ \ \\ 
\ \ (c)\ \ &\ \ 0.3\ \ &\ \ -1.2\ \ &\ \ 1\ \ &\ \ 1.571\ \ &\ \ 20.143\ \ &\ \ 0.967\ \ &\ \ -0.487\ \ \\
\ \ (d)\ \ &\ \ 0.3\ \ &\ \ -2\ \ &\ \ 1\ \ &\ \ 1.347\ \ &\ \ 26.21\ \ &\ \ 0.933\ \ &\ \ 0.861\ \ \\
\ \ (e)\ \ &\ \ 0.5\ \ &\ \ -2.3\ \ &\ \ 1\ \ &\ \ 1.82\ \ &\ \ 17.377\ \ &\ \ 0.879\ \ &\ \ -2.04\ \ \\
\ \ (f)\ \ &\ \ 0.5\ \ &\ \ -3.5\ \ &\ \ 1\ \ &\ \ 1.362\ \ &\ \ 21.47\ \ &\ \ 0.802\ \ &\ \ 1.082\ \ \\ [1ex]
\hline\hline
\end{tabular}
\caption{Parameters of the stable SPO around naked regularity.}
\label{nullnak1}
\end{table}

\begin{figure}%[H]
\centering
\includegraphics[scale=0.45]{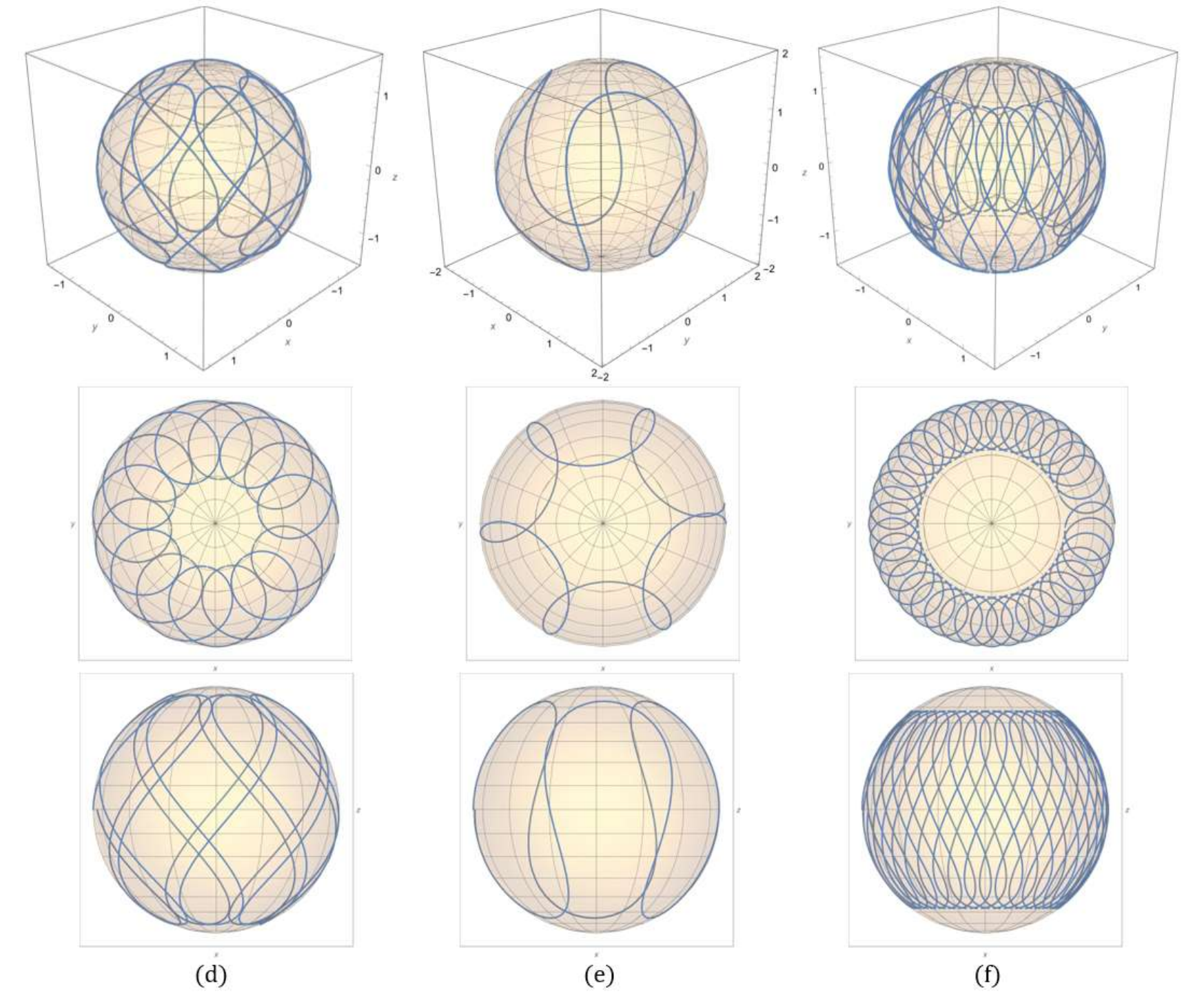}
\caption{Spherical photon orbits around no-horizon spacetime in the $(x-y-z)$ as well as the projective $x-y$ and $x-z$ planes. The direction of the black hole’s rotation is from west to east. (d) $\Phi=-2M$ and $Mk=0.3$, (e) $\Phi=-2.3M$ and $Mk=0.5$, (f) $\Phi=-3.5M$ and $Mk=0.5$.}
\label{fig:nak2}
\end{figure}

The absence of a horizon in the geometry opens avenues for stable SPOs (and STOs) by circumventing Eq.\eqref{socond2}. Here, we aim to explore the existence of stable spherical orbits around it, adhering to the condition:
\begin{equation}
\label{socond2stab}
\frac{d^2 R(r)}{dr^2}<0.
\end{equation}
However, our investigation uncovers that conditions~\eqref{socond2stab} and~\eqref{phic2} harmonize only when $\Phi\leq0$.
%The no-horizon geometry opens up the possibility of stable SPO (and STO) by evading Eq.~\eqref{socond2}. Here we try to investigate the {\it stable} spherical orbits around it, satisfying:
%\begin{equation}
%\label{socond2stab}
%\frac{d^2 R(r)}{dr^2}<0.
%\end{equation}  
%Our investigation, however, reveals that conditions~\eqref{socond2stab} and~\eqref{phic2} are compatible only when $\Phi\leq0$.
\begin{figure}%[H]
\centering
\includegraphics[scale=0.45]{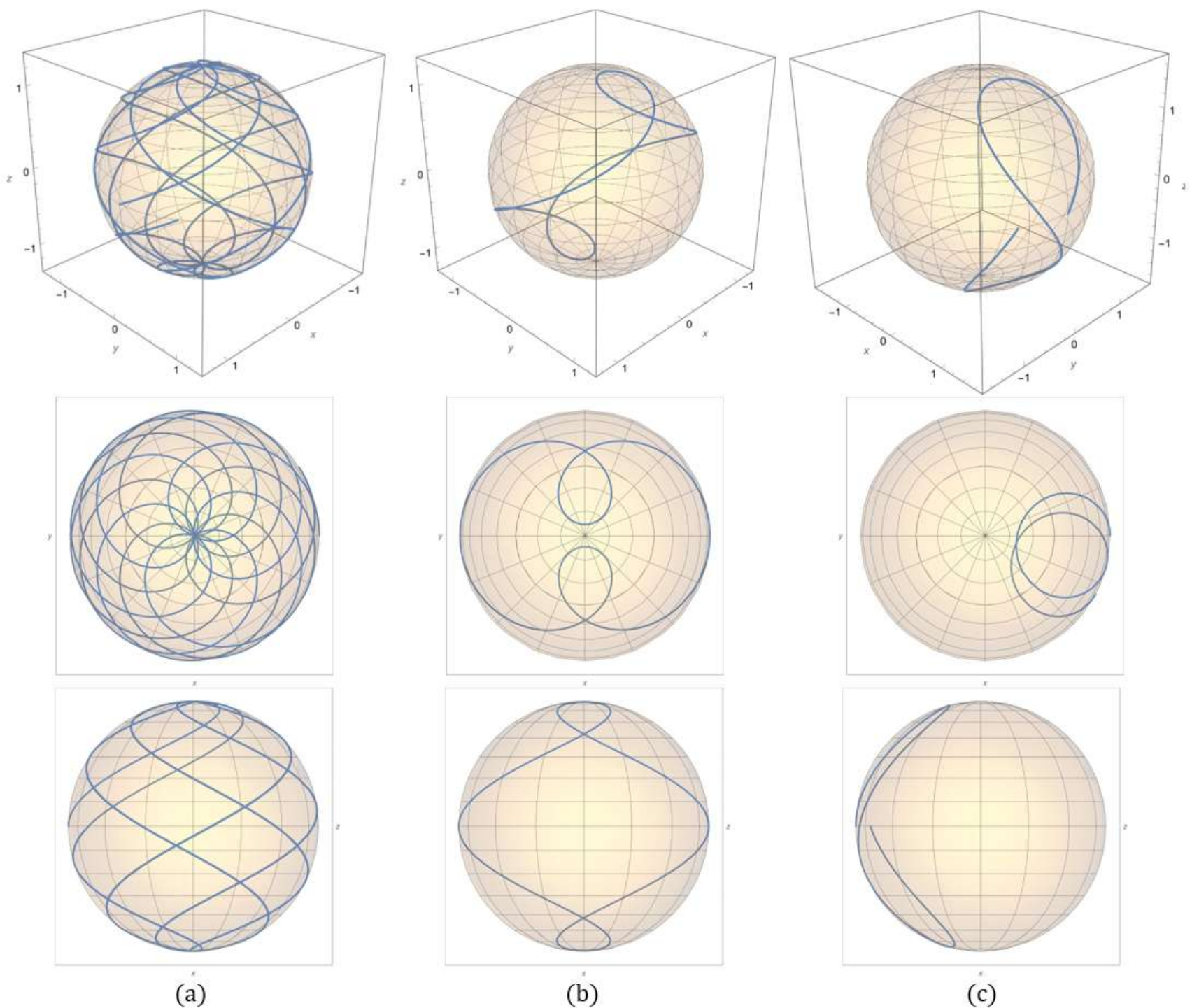}
\caption{Spherical photon orbits around no-horizon spacetime in the $(x-y-z)$ as well as the projective $x-y$ and $x-z$ planes. The direction of the black hole’s rotation is from west to east. (a) $\Phi=0$ and $Mk=0.1$, (b) $\Phi=-0.55 M$ and $Mk=0.1$, (c) $\Phi=-1.2 M$ and $Mk=0.3$.}
\label{fig:nak1}
\end{figure}
%We show the orbit trajectories around the no-horizon solution in Figs.~\ref{fig:nak1}-\ref{fig:nak2}, with parameters shown in Table~\ref{nullnak1}. We plot each figure for different values of $\Phi$ and $k$ because the requirement for stability places a rather stringent constraint on them. Since we have negative $\Phi$, the {\it Lens-Thirring effect} occurs. This effect has a greater impact on the equator and smaller on the poles. As can be seen, orbits with negative $\Phi$ values still have a prograde direction of motion around the equator, and only move in a retrograde direction when approaching the poles. This will be significant for orbits whose $\Phi$ value is not too far from 0 because the Lense-thirring effect has the power to offset the angular momentum of the photon.
We illustrate orbit trajectories around the no-horizon solution in Figs.~\ref{fig:nak1}-\ref{fig:nak2}, delineating parameters detailed in Table~\ref{nullnak1}. Each figure represents distinct values of $\Phi$ and $k$, as the stability requirement imposes stringent constraints on them. Given the negative $\Phi$ values, the occurrence of the {\it Lens-Thirring effect} is inevitable. This phenomenon exerts a pronounced influence on orbits near the equator compared to those near the poles. Notably, orbits with negative $\Phi$ values maintain a prograde motion around the equator, transitioning to a retrograde direction as they approach the poles. This observation holds particular significance for orbits with $\Phi\sim0$, as the Lens-Thirring effect possesses the capability to counteract the photon's angular momentum.

%\newpage

\subsection{Spherical Timelike Orbits}
\label{sec:STOR}

%\subsubsection{\bf Equation of motion}

The geodesics equations for an uncharged timelike particle are
\begin{eqnarray}
%\label{eq:optim}
%\begin{align}
\Sigma\dot{t}&=&-a^2 E \left(1-u^2\right)+\frac{1}{\Delta}\left[E\left(r^2+a^2\right)^2-2a\Phi m(r)r\right], \label{eq:cost}\\
\Sigma\dot{\phi}&=&\frac{\Phi}{1-u^2}+\frac{a}{\Delta}\left(2Em(r)r-a\Phi\right), \label{eq:const1}\\
\Sigma\dot{u}&=&\pm \sqrt{V\left(u\right)},\label{eq:const2}\\
\Sigma\dot{r}&=&\pm \sqrt{R\left(r\right)},\label{eq:const3}
%\end{align}
\end{eqnarray}
where
\begin{eqnarray}
\label{vutimelike}
R\left(r\right)&\equiv& \left(E^2-1\right)r^4+2m(r)r^3+ \left[a^2 \left(E^2-1\right)-\mathcal{C}-\Phi^2\right]r^2\nonumber\\
&&+2m(r)rr\left[\left(a E-\Phi\right)^2+\mathcal{C}\right]r  e^{-k/r} -a^2 \mathcal{C},\nonumber\\
V\left(u\right)&\equiv& a\left(1-E^2\right)u^4-\left[a^2\left(1-E^2\right)+\mathcal{C}+\Phi^2\right]+\mathcal{C}.\nonumber\\
%\Delta&=&r^2 +a^2 -2Mr e^{-\frac{k}{r}},\\
\end{eqnarray}

%\begin{eqnarray}
%%\label{eq:optim}
%%\begin{align}
%R\left(r\right)&=& \left(E^2-\mu^2\right)+2M\mu^2 r^3 e^{-\frac{k}{r}}+ \left[a^2 \left(E-\mu^2\right)-c-\Phi^2\right]r^2\\ \nonumber
%+2M\left[\left(a E-\Phi\right)^2+c\right]r  e^{-\frac{k}{r}} -a^2 c, \\
%    V\left(u\right) &= a\left(\mu^2-E^2\right)u^4-\left[a^2\left(\mu^2-E^2\right)+c+\Phi^2\right]+c, \label{vutimelike}\\
%    \Delta&=r^2 +a^2 -2Mr e^{-\frac{k}{r}}, \label{eq:const2} \\
%    \Sigma&= r^2+a^2\cos^2\theta \label{eq:const3}
%%\end{align}
%\end{eqnarray}
%The $\mu$ notation is rest mass of particle, for timelike particle is $\mu=1$.}

%\subsubsection{\bf Spherical Orbit Requirement}

As in the KN case above, for the timelike case here we only investigate the stable ones; that is when~\eqref{vutimelike} satisfies conditions~\eqref{socond} and~\eqref{socond2stab}. These conditions yield four types of solutions, the first two of which are
\begin{eqnarray}
E_{a,b} &=& \sqrt{ \frac{e^{-2k/r_{SO}} \left[-2a^4M \mathcal{C} e^{3k/r_{SO}} \left(k-r_{SO}\right) -A+r_{SO}^4 \left(B \pm 2 \sqrt{N}\right) \right] }{D}  },\nonumber\\
\Phi_{a,b} &=& -\frac{e^{-k/r_{SO}} E_{a,b}}{a H M (k-r_{SO})}\nonumber\\
&&\times\left[a^4 M^2 e^{2k/r_{SO}} (k-r_{SO})\left[k\left(c+r_{SO}^2\right)+r_{SO} \left(\mathcal{C}-r_{SO}^2\right)\right]+F+r_{SO}^6 \sqrt{\frac{G}{r_{SO}^8}}\right],\nonumber\\
\label{timelikereqreg}
\end{eqnarray}
where
\begin{eqnarray}
A &\equiv& a^2 e^{2k/r_{SO}} \bigg\{M^2 \left[r_{SO}^2 \left(k^2+5 \mathcal{C}\right)+k^2 \mathcal{C}-2 k \mathcal{C} r_{SO}+4 k r_{SO}^3-5 r_{SO}^4\right]\nonumber\\ 
&&M r_{SO}^3 e^{k/r_{SO}} [3 r_{SO} (r_{SO}-k)-4 \mathcal{C}]+\mathcal{C} r_{SO}^4 e^{2k/r_{SO}}\bigg\}, \\
B &\equiv& e^{k/r_{SO}} \left(r_{SO} e^{k/r_{SO}}-2 M\right)^2 \left[r_{SO} \left(r_{SO} e^{k/r_{SO}}-3 M\right)+k M\right],\\
N &\equiv& \frac{1}{r_{SO}^8} \bigg\{a^2 M^2 e^{3k/r_{SO}} (k-r_{SO})^2 \left[\left(a^2+r_{SO}^2\right) e^{k/r_{SO}}-2 M r_{SO}\right]^2\bigg\}\nonumber\\
&&\times\bigg\{\mathcal{C} e^{k/r_{SO}} \left(a^2 \mathcal{C}-r_{SO}^4\right)+M r_{SO}^2 \left[-k \left(\mathcal{C}+r_{SO}^2\right)+3 \mathcal{C} r_{SO}+r_{SO}^3\right]\bigg\},\\
D &\equiv& r_{SO}^4 \bigg\{4 a^2 M e^{k/r_{SO}} (k-r_{SO})+\left[r_{SO} \left(r_{SO} e^{k/r_{SO}}-3 M\right)+k M\right]^2\bigg\},\\
F &\equiv& a^2 M^2 r_{SO}^2 e^{k/r_{SO}} (k-r_{SO}) \bigg\{e^{k/r_{SO}} \left[k \left(\mathcal{C}+r_{SO}^2\right)-r_{SO} \left(\mathcal{C}+3 r_{SO}^2\right)\right]+4 M r_{SO}^2\bigg\},\\
G &\equiv& a^2 M^2 e^{3k/r_{SO}} (k-r_{SO})^2 \left[\left(a^2+r_{SO}^2\right) e^{k/r_{SO}}-2 M r_{SO}\right]^2\nonumber\\ 
&&\times\bigg\{\mathcal{C} e^{k/r_{SO}} \left(a^2 \mathcal{C}-r_{SO}^4\right)+M r_{SO}^2 \left[-k \left(\mathcal{C}+r_{SO}^2\right)+3 \mathcal{C} r_{SO}+r_{SO}^3\right]\bigg\},\\
H &\equiv& a^2 e^{k/r_{SO}} \left[r_{SO} \left(\mathcal{C} r_{SO} e^{k/r_{SO}}-M \mathcal{C}+M r_{SO}^2\right)-k M \left(\mathcal{C}+r_{SO}^2\right)\right]-r_{SO}^4 \left(r_{SO} e^{k/r_{SO}}-2 M\right)^2.\nonumber\\
\end{eqnarray}
The third and fourth solutions are related through 
\begin{equation}
    (E_{c,d},\Phi_{c,d})=-(E_{a,b},\Phi_{a,b}).
\end{equation}

\subsubsection{\bf Analytical Solutions}

From $V(u^2)=0$ the roots $w_{1,2}\equiv u^2$ can be found as
\begin{equation}
    w_{1,2} = \frac{a^2 \left(1-E^2\right)+\mathcal{C}+\Phi ^2 \mp \sqrt{\left[a^2 \left(1-E^2\right)+\mathcal{C}+\Phi ^2\right]^2-4 a^2 \mathcal{C} \left(1-E^2\right)} }{2a^2 \left(1-E^2\right)}. 
\end{equation}
%{\it Bound Orbit}
As before, for bound orbits we have $E^2<1$ and $0 \leq w \leq w_1 \leq 1 < w_2$. The analytical solutions for $\left\{\theta(\lambda),\phi(\lambda),t(\lambda)\right\}$ are 
\begin{eqnarray}
\theta(\lambda)&=& \arccos\left[\sqrt{w_1}\ \text{sn}(a \sqrt{\left(1-E^2\right) w_2}\lambda,\kappa)\right],\label{thetime} \\
\phi(\lambda)&=& \frac{\Phi  }{a \sqrt{\left(1-E^2\right) w_2}} \Pi (\Psi,w_1,\kappa) + \frac{a}{\Delta} \left(2MrE e^{-k/r} - a\Phi \right)\lambda,\label{phitime} \\
t(\lambda)&=& - \frac{a E}{\sqrt{(1-E^2)w_2}}\left[ (1-w_2) F(\Psi,\kappa) + w_2 E(\Psi,\kappa)\right] \nonumber\\
&&+ \frac{1}{\Delta}\left[ E(r^2+a^2)^2 - 2Mra \Phi e^{-k/r} \right] \lambda,\label{ttime} 
\end{eqnarray}
where %sn is Jacobi sn function and $E(\psi,k)$ is the second kind of incomplete elliptic integral, and
\begin{eqnarray}
\Psi &\equiv& \text{am}(a \sqrt{\left(1-E^2\right) w_2}\lambda,\kappa), \\ \nonumber
\kappa &\equiv& \sqrt{\frac{w_1}{w_2}}.
\end{eqnarray}
The period of $\theta(\lambda)$ is
\begin{equation}
\Delta \lambda = \frac{4}{a \sqrt{\left(1-E^2\right) w_2}}K(\kappa).
\end{equation}
The change in $\phi$ and $t$ for one period are
\begin{eqnarray}
\Delta \phi &=& \frac{4}{a \sqrt{\left(1-E^2\right) w_2}} \left[ \frac{\Phi}{a} \Pi(w_1,\kappa) + \frac{2MrE e^{-k/r} - a\Phi}{\Delta} K(\kappa)\right],\\ 
\Delta t &=&-\frac{4}{a \sqrt{\left(1-E^2\right) w_2}} \bigg\{ -a\left[(1-w_2) K(\kappa) + w_2 E(\kappa)\right]\nonumber\\ 
&&+ \left[ \frac{E(r^2+a^2)^2 - 2Mra \Phi e^{-k/r}}{a \Delta} \right]K(\kappa)\bigg\}.
\end{eqnarray}

%{\it Unbound Orbit}

%In this case, we consider that $E^2>1$ so the range of $w$ are $w_2 < 0 \leq w \leq w_1 \leq 1$. The solution for this case in $(\theta,\phi,t)$ are

%\begin{eqnarray}
%    \theta &=& \arccos{(\sqrt{-w_2} \kappa \text{sd}(a \sqrt{\left(E^2-1\right) (w_1-w_2)}\lambda,\kappa))}, \\
%    \phi &=& \frac{\Phi}{a(1-w_2)\sqrt{(E^2-1)(w_1-w_2)}} [ F(\Psi,\kappa) - w_2 \Pi(\Psi,\kappa^2(1-w_2),\kappa) ] \\ \nonumber
%    &+& \frac{a}{\Delta} \left(2MrE e^{-\frac{k}{r}} - a\Phi \right)\lambda, \\ 
%    t &=& - \frac{aE}{\sqrt{(E^2-1)(w_1-w_2)}} [ (1-w_2) F(\Psi,\kappa) + w_2 \Pi(\Psi,\kappa^2,\kappa) ] \\ \nonumber
%    &+& \frac{1}{\Delta}\left[ E(r^2+a^2)^2 - 2Mra \Phi e^{-\frac{k}{r}} \right] \lambda, 
%\end{eqnarray}
%where
%\begin{eqnarray}
%\Psi &\equiv& \text{am}(a \sqrt{\left(E^2-1\right) (w_1-w_2)}\lambda,\kappa), \\ \nonumber
%\kappa &\equiv& \sqrt{\frac{w_1}{w_1-w_2}}.
%\end{eqnarray}
%$\theta$ is a periodic function of $\lambda$, with period
%\begin{equation}
%    \Delta \lambda = \frac{4}{a \sqrt{\left(E^2-1\right) (w_1-w_2)}}K(\kappa).
%\end{equation}
%The change in $\phi$ and $t$ for one period are 
%\begin{eqnarray}
%    \Delta \phi &=& \frac{4}{\sqrt{(E^2-1)(w_1-w_2)}} \left[ \frac{\Phi}{a(1-w_1)} \Pi(-\frac{-w_1}{1 - w_1},\kappa) + \frac{2MrE e^{-\frac{k}{r}} - a\Phi}{\Delta} K(\kappa) \right],  \\
%    \Delta t &=& \frac{4aE[(1-w_2) K(\kappa) + w_2 \Pi(\kappa^2,\kappa)]}{\sqrt{(E^2-1)(w_1-w_2)}} -4 \left[ \frac{E(r^2+a^2)^2 - 2Mra \Phi e^{-\frac{k}{r}}}{a \Delta\sqrt{(E^2-1)(w_1-w_2)}} \right]K(\kappa) .\nonumber \\
%\end{eqnarray}

\subsubsection{\bf Selected orbits}

\begin{figure}%[H]
\centering
\includegraphics[scale=0.53]{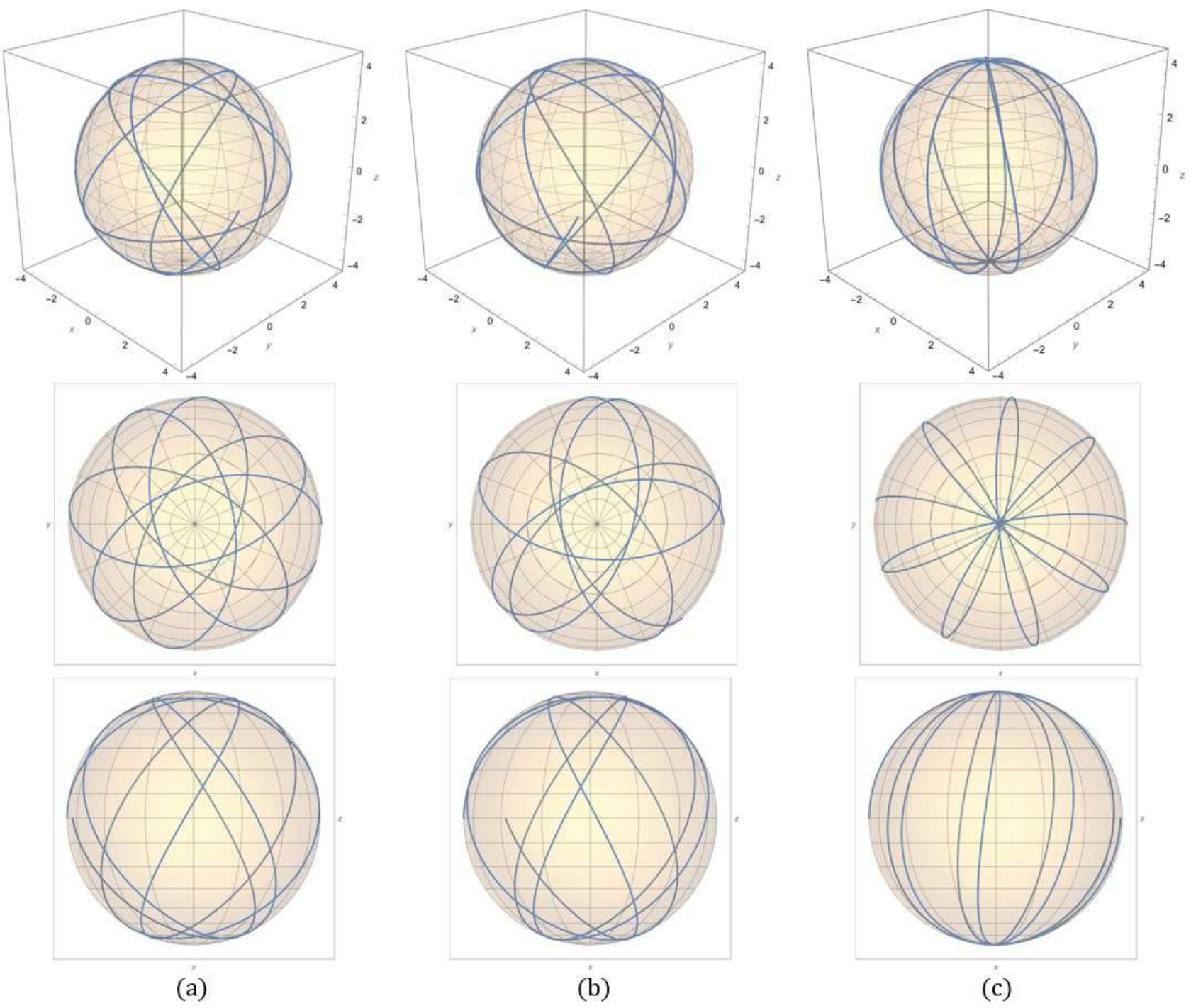}
\caption{Spherical timelike orbits around extremal regular black hole in the $(x-y-z)$ as well as the projective $x-y$ and $x-z$ planes. The direction of the black hole’s rotation is from west to east. (a) $\Phi/M=0.918$ and $k=0$, (b) 
$\Phi/M=0.819$ and $k=0.1$, (c) $\Phi/M=0.043$ and $k=0.5$.}
\label{fig:Phi1}
\end{figure}

\begin{figure}%[H]
\centering
\includegraphics[scale=0.45]{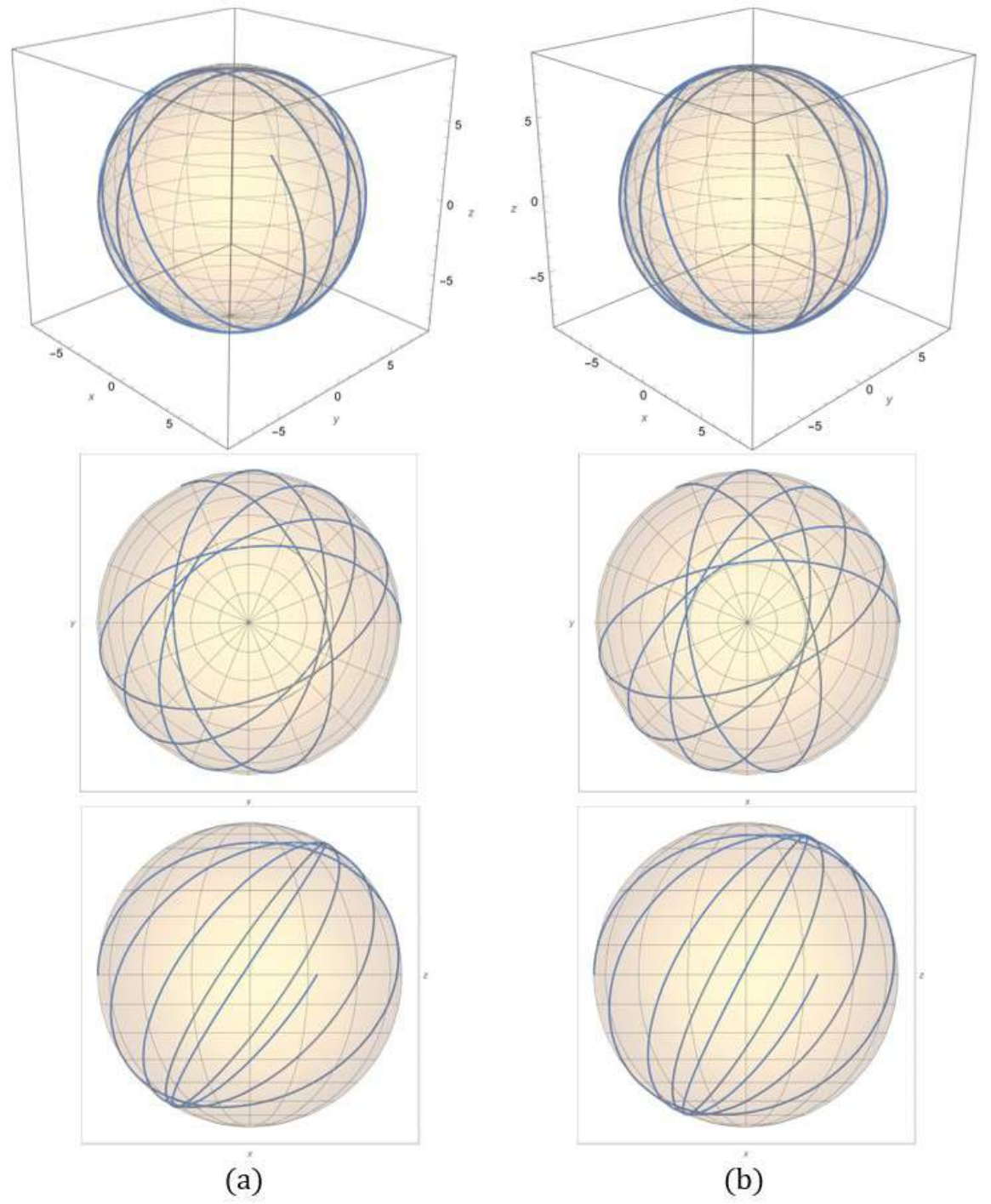}
\caption{Spherical timelike orbits around no-horizon spacetime in the $(x-y-z)$ as well as the projective $x-y$ and $x-z$ planes. The direction of the black hole’s rotation is from west to east. (a) $\Phi/M=1.596$ and $k=0.1$, (b) $\Phi/M=1.211$ and $k=0.5$.}
        \label{fig:Phi2}
\end{figure}

From the derived solutions~\eqref{thetime}-\eqref{ttime}, we can visualize certain timelike orbits.%From the solutions~\eqref{thetime}-\eqref{ttime} we can plot some timelike orbits.% We are now in a position to plot the orbits.% In the timelike case, the only parameter we change is the value of $k$ to see if the $k$ factor that causes this irregularity has an impact on the trajectory of the timelike orbit. Then keep in mind that in the timelike case, it is possible to obtain a stable orbit in each condition. The example orbits below are stable orbits.
%\\
%\\
%{\it Non Extremal Orbit}
\begin{table}% [H]
\centering
\begin{tabular}{||c c c c c c c c c||} 
 \hline\hline
\ \ Orbit\ \ &\ \ $Mk$\ \ &\ \ $a/M$\ \ &\ \ $r_{SO}/M$\ \ &\ \ $C/M^2$\ \ &\ \ $E$\ \ &\ \ $\Phi$/M\ \ &\ \ $u_0$\ \ &\ \ $\Delta\phi$\ \ \\ [0.5ex]
 \hline
\ \ (a)\ \ &\ \ 0\ \ &\ \ 0.3\ \ &\ \ 6\ \ &\ \ 8\ \ &\ \ 0.937\ \ &\ \ 1.733\ \ &\ \ 0.852\ \ &\ \ 6.535\ \ \\ 
\ \ (b)\ \ &\ \ 0.1\ \ &\ \ 0.3\ \ &\ \ 6\ \ &\ \ 8\ \ &\ \ 0.936\ \ &\ \ 1.576\ \ &\ \ 0.873\ \ &\ \ 6.535\ \ \\
\ \ (c)\ \ &\ \ 0.5\ \ &\ \ 0.3\ \ &\ \ 6\ \ &\ \ 8\ \ &\ \ 0.93\ \ &\ \ 0.857\ \ &\ \ 0.956\ \ &\ \ 6.537\ \ \\
 \hline\hline
\end{tabular}
\caption{Parameters of the STO around a non-extremal regular rotating black hole.}
\label{timeno}
\end{table}
\begin{figure}%[H]
         \centering
\includegraphics[scale=0.45]{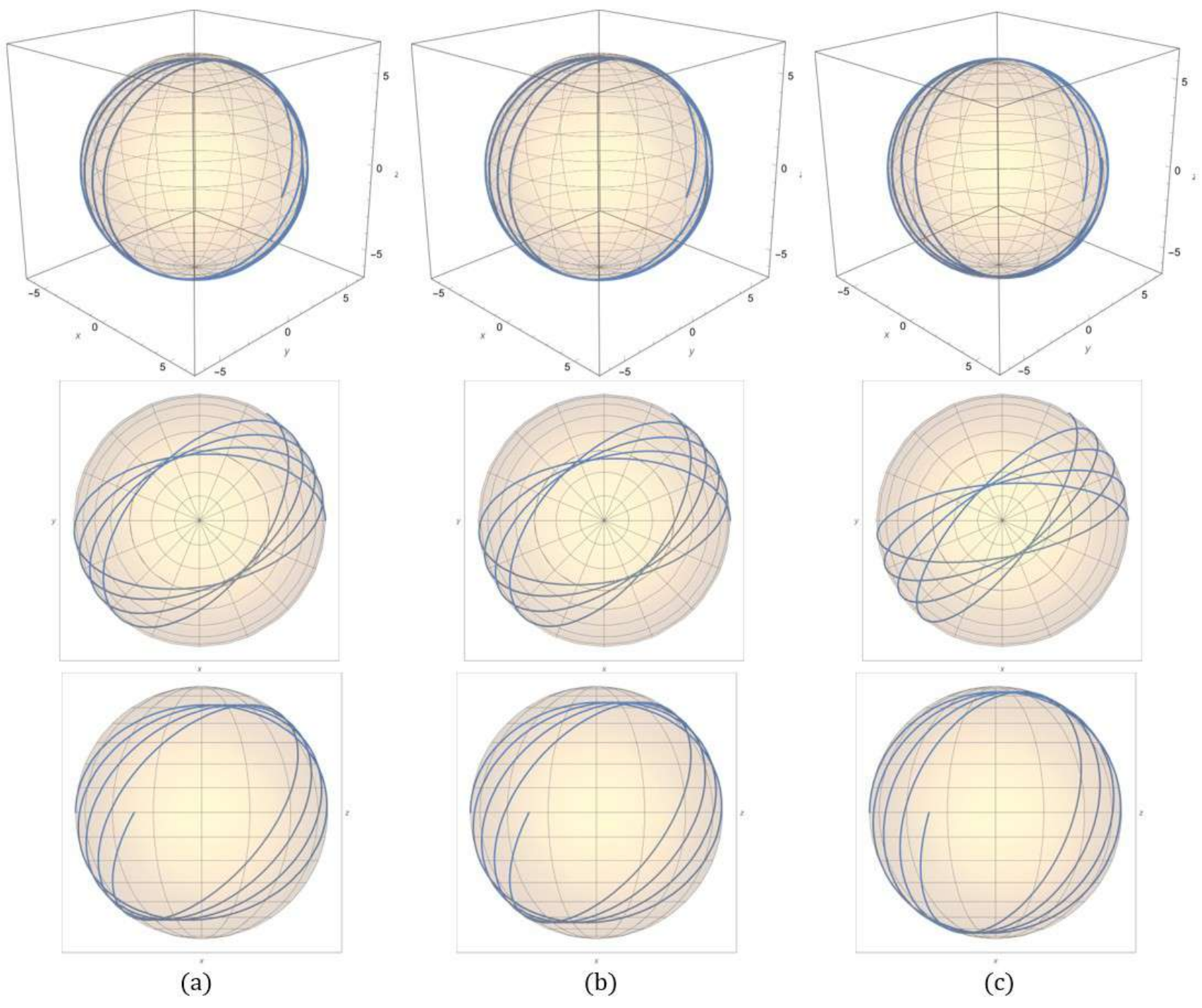}
\caption{Spherical timelike orbits around non-extremal regular black hole case in the $(x-y-z)$ as well as the projective $x-y$ and $x-z$ planes. The direction of the black hole’s rotation is from west to east. (a) $\Phi/M=1.733$ and $Mk=0$, (b) $\Phi/M=1.576$ and $Mk=0.1$, 
(c) $\Phi/M=0.857$ and $Mk=0.5$.}
\label{fig:Phi0}
\end{figure}
We present the STOs for non-extremal cases in Fig.~\ref{fig:Phi0}, accompanied by parameters detailed in Table~\ref{timeno}. Analogous to the Kerr-Newman (KN) and regular SPO scenarios, there is an inclination for larger values of $k$ to draw the orbit trajectory nearer to the pole, resulting in a notably expanded latitudinal oscillation amplitude. Distinctive plots for STOs encircling extremal black holes are showcased in Fig.~\ref{fig:Phi1}, with corresponding parameters outlined in Table~\ref{timeex}. Here, we observe a distinctive orbit morphology: with increasing $k$, the trajectories depicted in the projective $x-y$ plane exhibit diminishing "leaf" shapes and augmented latitudinal oscillation amplitudes. Lastly, we delineate the orbits surrounding naked or no-horizon scenarios in Fig.~\ref{fig:Phi2}, accompanied by parameters expounded in Table~\ref{timenak}.%We illustrate the STOs for {\it }non-extremal cases in Fig.~\ref{fig:Phi0} with parameters shown in Table \ref{timeno}. As in the KN as well as the regular SPO cases, there is a tendency that the larger $k$ causes the orbit trajectory to get closer to the pole, or practically have a larger latitudinal oscillation amplitude. Typical plots for the STO around the {\i textremal} black holes are shown in Fig.~\ref{fig:Phi1} with parameters shown in Table \ref{timeex}. Here we see a more unique orbit shape. As $k$ increases the trajectories seen from the projective $x-y$ plane show smaller "leaf" shapes and larger latitudinal oscillation amplitudes. Finally, we illustrate the orbits around the {\it naked/no-horizon} cases in Fig.~\ref{fig:Phi2} with parameters shown in Table \ref{timenak}. %As in the KN as well as the regular SPO cases, we observe that increasing $k$ causes the orbit to have a larger latitudinal oscillation amplitude.
\begin{table}% [H]
\centering
\begin{tabular}{||c c c c c c c c c||} 
 \hline\hline
\ \ Orbit\ \ &\ \ $Mk$\ \ &\ \ $a/M$\ \ &\ \ $r_{SO}/M$\ \ &\ \ $C/M^2$\ \ &\ \ $E$\ \ &\ \ $\Phi/M$\ \ &\ \ $u_0$\ \ &\ \ $\Delta\phi$\ \ \\ [0.5ex]
 \hline
\ \ (a)\ \ &\ \ 0\ \ &\ \ 1\ \ &\ \ 4\ \ &\ \ 8\ \ &\ \ 0.918\ \ &\ \ 0.918\ \ &\ \ 0.95\ \ &\ \ 7.78\ \ \\ 
\ \ (b)\ \ &\ \ 0.1\ \ &\ \ 0.8998\ \ &\ \ 4\ \ &\ \ 8\ \ &\ \ 0.918\ \ &\ \ 0.819\ \ &\ \ 0.96\ \ &\ \ 7.645\ \ \\
\ \ (c)\ \ &\ \ 0.5\ \ &\ \ 0.472\ \ &\ \ 4\ \ &\ \ 8\ \ &\ \ 0.918\ \ &\ \ 0.043\ \ &\ \ 0.999\ \ &\ \ 7.023\ \ \\
\hline\hline
\end{tabular}
\caption{Parameters of the STO around the extremal regular rotating black hole.}
\label{timeex}
\end{table}

%We investigate the shape of $V_{eff}$ for several values of $\Phi$ in this extremal case. The result is shown in Fig.~\ref{fig:regular isso}. It can be seen that as $\Phi$ increases the extrema flattens out and the trend goes monotonically down. However, no matter how much $\Phi$ increases there always exist two extremes, one unstable inside the horizon and one stable exactly at it. This {\it innermost spherical orbit} is thus unstable, located at $r=0.04M$.
%\\
%\begin{figure}%[H]
%\centering
%\includegraphics[scale=0.7]{reguler ISSO.pdf}
%\caption{$V_{eff}$ extremal Ghosh for various $\Phi$. The thick, dashed, and dotted-dashed are $\Phi=0.25M$, $\Phi=0.819M$, and $\Phi=1.219M$ respectively. The dotted vertical line indicates the extremal horizon. Here we set $E=0.941$ and other parameters from Table~\ref{timeex} (b). There is an innermost unstable spherical orbit at $r=0.04M$.}
%        \label{fig:regular isso}
%\end{figure}

%{\it Naked Orbits}
\begin{table}% [H]
\centering
\begin{tabular}{||c c c c c c c c c||} 
 \hline\hline
\ \ Orbit\ \ &\ \ $Mk$\ \ &\ \ $a/M$\ \ &\ \ $r/M$\ \ &\ \ $C/M^2$\ \ &\ \ $E$\ \ &\ \ $\Phi/M$\ \ &\ \ $u_0$\ \ &\ \ $\Delta\phi$\ \ \\ [0.5ex]
 \hline
\ \ (a)\ \ &\ \ 0.1\ \ &\ \ 1\ \ &\ \ 8\ \ &\ \ 8\ \ &\ \ 0.943\ \ &\ \ 1.596\ \ &\ \ 0.869\ \ &\ \ 6.79\ \ \\ 
\ \ (b)\ \ &\ \ 0.5\ \ &\ \ 1\ \ &\ \ 8\ \ &\ \ 8\ \ &\ \ 0.941\ \ &\ \ 1.211\ \ &\ \ 0.918\ \ &\ \ 6.798\ \ \\
 \hline\hline
\end{tabular}
\caption{Parameters of the STO around naked regularity.}
\label{timenak}
\end{table}

 %Similar to the extremal case, ISSO does not exist in naked regularity. The shape of the effective potential is shown in Fig.~\ref{fig:naked isso}, where there is the innermost unstable spherical orbit located at $r=0.25M$. The corresponding orbit is planar, not spherical.

%\begin{figure}%[H]
%\centering
%\includegraphics[scale=0.7]{naked isso.pdf}
%\caption{$V_{eff}$ of Ghosh's naked regularity for various $\Phi$. Here the thick, dashed, and the dotted-dashed lines are for $\Phi=0.25M$, $\Phi=0.8998M$, and $\Phi=1.255M$,  respectively. Here we follow parameters from the Table~\ref{timenak} (b). There is an innermost unstable point at $r=0.145M$.}
%        \label{fig:naked isso}
%\end{figure}

%\newpage

\subsubsection{\bf Innermost Stable Spherical Orbits}

Using a similar method with Kerr-Newman black holes, we can get the ISSO as a function of Carter's constant. The ISSO for the Ghosh BH presents some interesting features: the branches are discontinuous. Some of them stop at some finite values, as we shall see below. The form of the regular metric is such that there exist some ranges of $C/M^2$ which have complex $r_{ISSO}$.

\begin{figure}%[H]
         \centering
\includegraphics[scale=0.6]{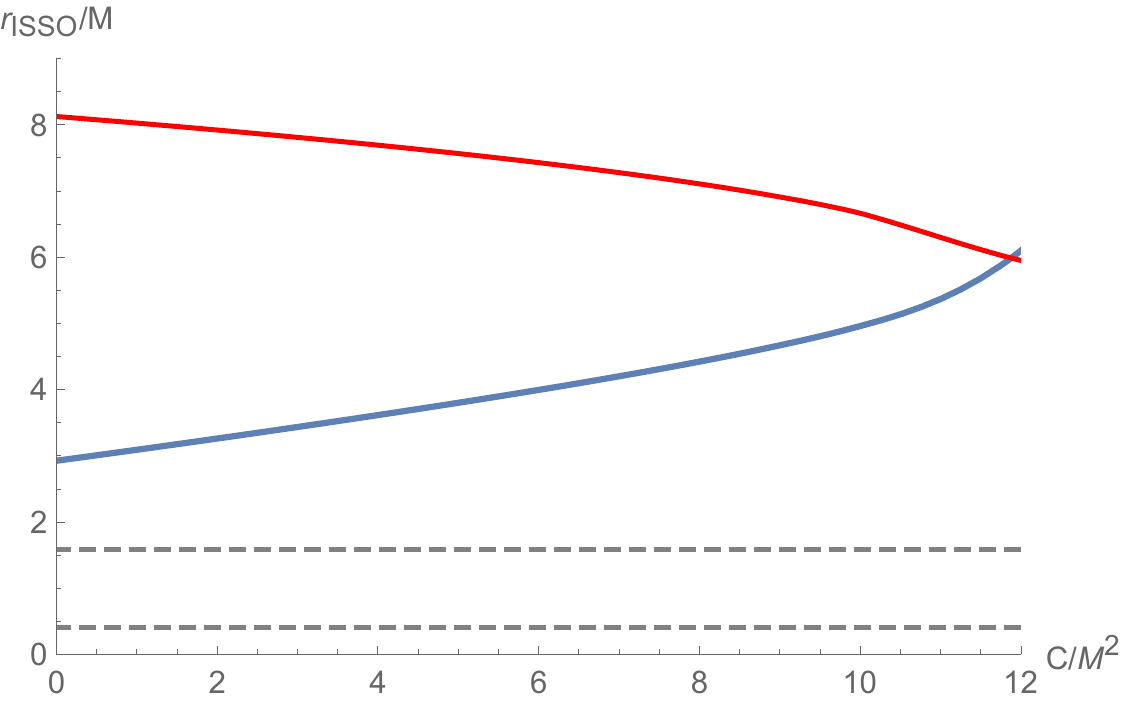}
\caption{$r_{ISSO}$ as a function of Carter's constant for a non-extremal regular black hole when $a=0.7M$ and $k=0.1M$. The red and blue lines are the $r_{isso}$ from the $\pm$ sign in eq. \ref{timelikereqreg}, and black dashed lines are the event horizons.}
\label{fig:issononex}
\end{figure}

\begin{figure}%[H]
         \centering
\includegraphics[scale=0.6]{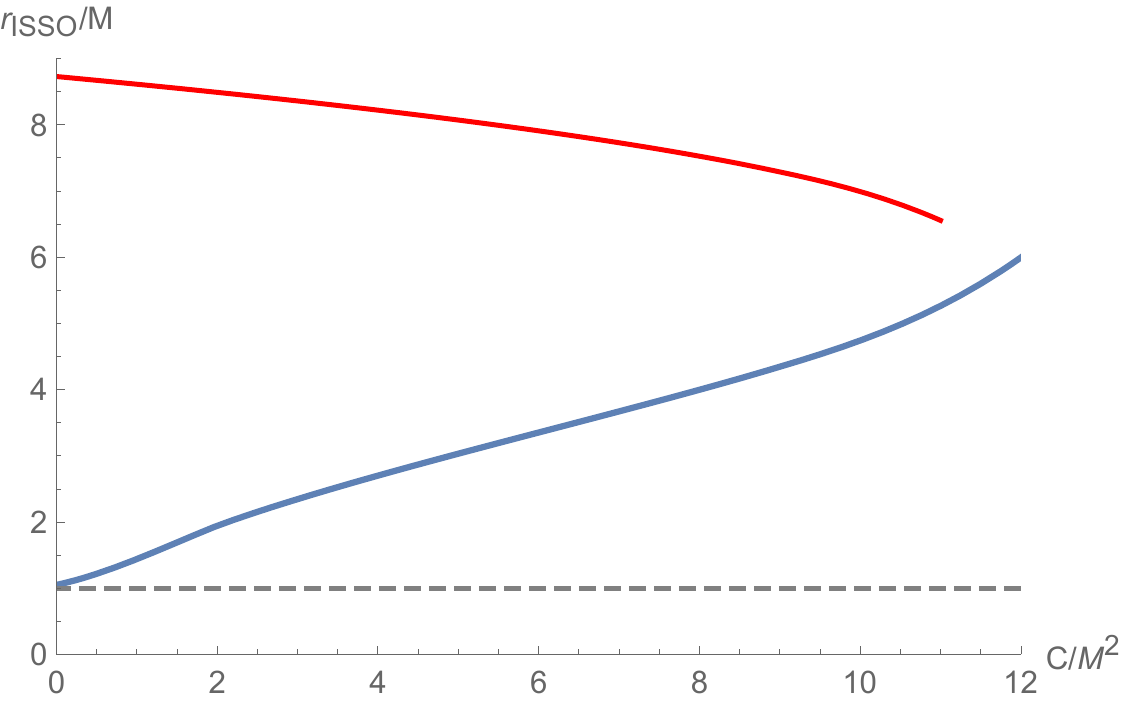}
\caption{$r_{ISSO}$ as a function of Carter's constant for an extremal regular black hole when $a=0.8998M$ and $k=0.1M$. The red and blue lines are the $r_{isso}$ from $\pm$ sign in eq. \ref{timelikereqreg}, and the black dashed line is the event horizon.}
\label{fig:issoex}
\end{figure}

\begin{figure}%[H]
         \centering
\includegraphics[scale=0.6]{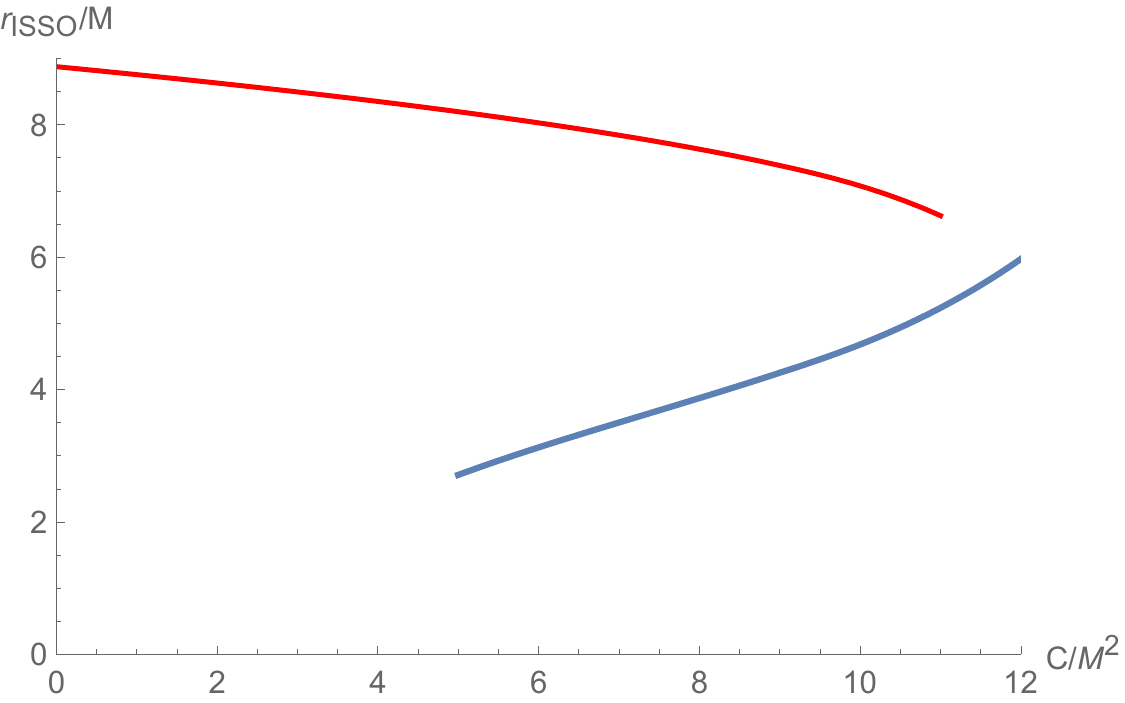}
\caption{$r_{ISSO}$ as a function of Carter's constant for a naked regular black hole when $a=0.95M$ and $k=0.1M$. The red and blue lines are the $r_{isso}$ from the $\pm$ sign in eq. \ref{timelikereqreg}.}
\label{fig:issoreg}
\end{figure}

\begin{figure}%[H]
         \centering
\includegraphics[scale=0.43]{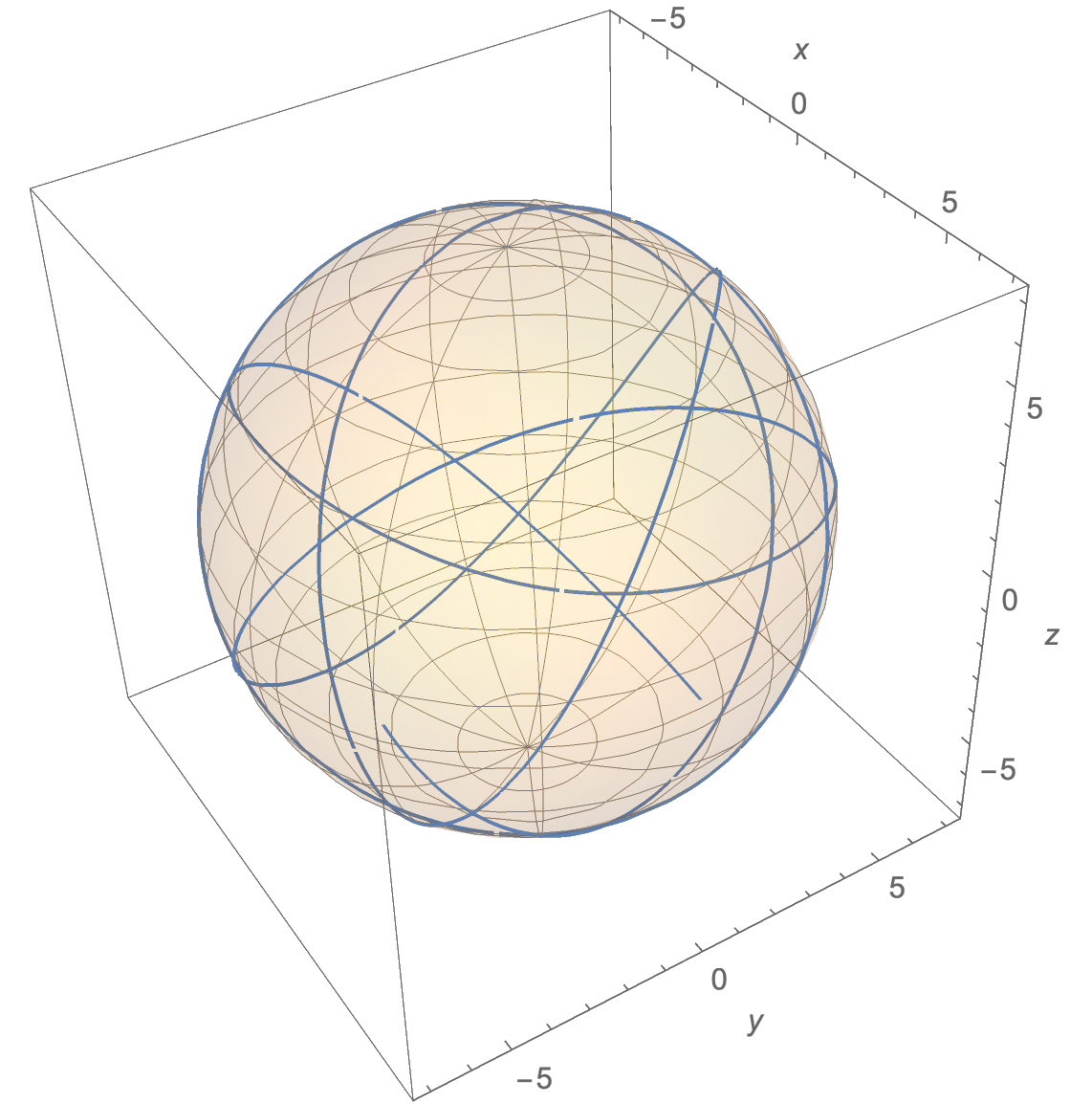}
\caption{The ISSO trajectory around the no-horizon case of Ghosh BH (based on the data in Fig.~\ref{fig:issoreg}). Here $C/M^2=10$, $a/M=0.95$, and $k/M=0.1$. The corresponding ISSO radius is $r_{ISSO}\sim7.075M$.}
\label{fig:issotrajectory}
\end{figure}

For a non-extremal case the result is shown in Fig.~\ref{fig:issononex} with $a=0.7M$ and $k=0.1M$. Both branches are observable (outside the horizons). The lower branch starts at $r_{ISSO}/M\sim2.92$ while the upper one starts $r_{ISSO}/M\sim8.12$. Unlike the KN case, here the two branches do not merge and stop at $C_{crit}$, but they somehow cross as can be seen from the figure. The crossing value is around $r_{ISSO}/M\sim6.09$ at $C/M^2\sim12$. A more interesting case happens for the extremal BH. In Fig.~\ref{fig:issoex} it can be seen that the lower branch starts at the extremal horizon for $C/M^2=0$, then goes up as usual. However, instead of merging or crossing with the upper branch, the two branches do not merge at a single $C$ value. The upper branch stops being real at around $r_{ISSO}/M\sim6.55$. It only exists between $0\leq C/M^2\lesssim11$. Finally, the no-horizon case is the most interesting of all. As can be seen from Fig.~\ref{fig:issoreg}, the lower branch ceases to exist up until $C/M^2\sim4.4$, at $r_{ISSO}\sim2.23$. On the other hand, the upper branch starts with $r_{ISSO}/M\sim8.8$ and it stops being real around $r_{ISSO}\lesssim5.98$ at $C/M^2\sim11$. A typical ISSO trajectory is shown in Fig.~\ref{fig:issotrajectory}. This is an example of ISSO in a no-horizon case, with $a/M=0.95$, $k/M=0.1$, and $C/M^2=10$. The corresponding ISSO radius is $r_{ISSO}/M\sim7.08$.

%\textcolor{red}{Furtheermore, we can get the trajectory of ISSO in Fig. \ref{fig:issotrajectory}}

%\textcolor{blue}{Furthermore, we can get the ISSO as a function of Carter's constant for an extremal case in Fig.~\ref{fig:issoex}. In this case, the red lines ISSO only exist between $C=0$ and $C= 11M^2$.}

\section{Conclusion}
\label{sec:conc}

In this study we expand upon Teo's seminal analysis of spherical orbits~\cite{teo, Teo:2020sey}, delving into the charged rotating black holes. Specifically, our focus lies on two distinct models: the Kerr-Newman and the Ghosh black holes, the latter representing a novel regular rotating black hole obtained by conducting the Newman-Janis transformation of a spherically-symmetric regular metric. Notably, the metric described by Eq.~\eqref{regrot} reduces to Kerr at $k\rightarrow0$ and to the Kerr-Newman at $r>>k$. Leveraging exact analytical solutions in terms of Mino parameters, we explore the presence of both spherical photon orbits (SPOs) and spherical timelike orbits (STOs) in these models. Noteworthy is our assumption that the direction of rotation for the black hole is from west to east, with all orbits originating from the equatorial plane and progressing northward.

%In this work, we extend Teo's analysis of spherical orbits~\cite{teo, Teo:2020sey} to the case of charged rotating black holes. In particular, we study two rotating models: the Kerr-Newman and the Ghosh (regular) black holes. The latter can be perceived as the generalization of charged black holes sourced by NLED charge. The metric~\eqref{regrot} reduces to Kerr at $k\rightarrow0$ and Kerr-Newman at $r>>k$. The exact analytical solutions for the geodesics can be expressed in terms of the Mino parameters. We investigate the existence of SPOs and STOs in both models. All orbits begin from the equator and head northwards. In this work we assume that the direction of the black hole's rotation is from west to east.%For regular black hole, the naked condition does not lead to the exposure of singularity by an external observer. Rather, it exposes the interior de Sitter-like space. We dubbed it {\it the naked regularity}.  

In the case of Kerr-Newman black holes we found that, similar to the Kerr case, all observable SPOs prove to be unstable under radial perturbation. This instability is evident from the effective potential 
$V_{eff}\equiv-R(r)$, where all SPOs satisfy Eq.~\eqref{socond2}. With the introduction of charge, the $C-r$ and $\Phi-r$ plots undergo a shift to the right, while the maximum value of the former decreases compared to Kerr. This shift indicates a higher concentration of photon trajectories around the equatorial plane. This concentration is further confirmed by the directional velocity $\theta$ and angular momentum $\Phi$ which decrease and increase, respectively, as the charge 
$Q$ increases. Additionally, it requires more latitudinal oscillations to form a closed trajectory. Notably, the charge 
$Q$ can reorient the photon's motion about the z-axis of the black hole. For $\Phi=-M$, the initially retrograde orbit transitions to prograde motion. This transition occurs at 
$Q/M=0.6$ for both non-extremal and extremal black holes. Furthermore, we observe a photon boomerang trajectory, $\Delta\phi=\pi$, in the extremal case. This happens in the extremal case, occurring at low charge values $Q=0.02528M$, and $\Phi=0$. Here we do not attempt to investigate the stable SPOs, which are available only for Kerr naked singularity. Note that there has been extensive discussions on the SPOs around the Kerr naked singularities (KNSs)~\cite{Charbulak:2018wzb, Charbulak:2017bpj, Nguyen:2023clb}. The existence of Kerr superspinars might be a signature for the string theory~\cite{Gimon:2007ur}.

Exploring geodesics around Ghosh black holes raises the intriguing question of whether stable SPOs exist outside the horizon. This is a nonsingular rotating black hole, parametrized by an additional parameter $k$. However, our investigation yielded disappointing results, as all SPOs around regular black holes prove to be unstable. Additionally, we observed that the condition 
$Mk>>1$ is incompatible with both non-zero $a/M$ and observability criteria $r_{SO}>r_h$. In the non-extremal case, this circumstance renders photon orbits predominantly equatorial. However, in the extremal case, where 
$Mk\lesssim0.5$, a complete SPO exists. Note that in this BH the extremality can be achieved by either $a/M$ or $Mk$. We identified an approximate photon boomerang in this extremal case, with $k=0.05$ and $\Phi=-0.07$.
Since regular black holes lack naked singularities, the absence of a horizon in the metric poses no physical paradox. This so-called "{\it no-horizon}" condition may conceptually align with that of a horizonless ultracompact star. Digging further into our investigation around no-horizon condition, we discovered that photons can orbit spherically under stable conditions, albeit only when $\Phi\leq0$. This finding highlights the presence of the Lens-Thirring effect.

%One motivation for studying geodesics around the regular black hole is whether there exist stable SPOs outside the horizon. It appears that our attempt was futile. All SPOs around regular black holes are also unstable. We also found that $Mk>>1$ is not compatible with both non-zero $a/M$ and conditions for observability ($r_{SO}>r_h$). For the non-extremal case, this situation makes the photon orbits almost trivially equatorial. The situation is a little bit better for the extremal case, where there exists full SPO for $Mk\lesssim0.5$. We obtain an approximate photon boomerang in this extremal case, with $k=0.05$ and $\Phi=-0.07$. Since naked singularity is absent in regular black holes, there is no physical problem should the metric be devoid of any horizon. This ``{\it no-horizon} condition" might physically be related to a horizonless ultracompact star. Our investigation around naked regularity reveals that photons can orbit spherically in {\it stable} condition, but this is possible only when $\Phi\leq0$. This results in the existence of the Lens-Thirring effect. 

For stable spherical timelike orbits (STOs), both KN and Ghosh BHs allow orbits to exist outside their horizons. In the KN scenario, bound orbits occur when 
 $E^2<1$, while $E^2=1$ signifies a marginally bound orbit. Unlike null orbits, timelike orbits do not exhibit retrograde-prograde switching. In the Ghosh case, higher values of $k$ drive trajectories closer to the pole, resulting in larger latitudinal oscillation amplitudes. Notably, the concept of innermost stable spherical orbits (ISSOs) in spherical trajectory parallels the innermost stable circular orbits (ISCOs) in equitorial motion. The ISSO radius in KN can be determined by solving Eqs.~\eqref{KNisso1}-\eqref{KNisso2}. These solutions, depicted in Fig.~\ref{fig:rissokn}, manifest as two branches of $r_{ISSO}\left(C\right)$. A typical ISSO trajectory in the KN BH, illustrated in Fig.~\ref{fig:pi16}, appears densely concentrated around the equator initially, gradually filling the entire solid angle as it completes one latitudinal oscillation. In contrast, ISSO behavior in the Ghosh BH presents intriguing nuances. Unlike the KN case, the nonsingular nature of the Ghosh metric causes ISSO branches to terminate at finite values, as depicted in Figs.~\ref{fig:issononex}-\ref{fig:issoreg}. A representative ISSO trajectory is shown in Fig.~\ref{fig:issotrajectory}. As a closing statement in this paragraph we should briefly comment on this ISSO issue. It is evident that in both types of black holes under examination, the radial potential $R(r)$ is at least quartic in $r$. This leads to two roots for each solution of $R''=0$, hence the presence of two branches. The lower branch in each 
$r_{ISSO}$ vs $C$ plot should be the one that represents the "true" ISSO, as the name by itself stands.

Finally, we should also comment on the phenomenological aspects of our results. Our theoretical findings hold significant potential for astrophysical applications. 
Firstly, the prevalence of unstable spherical photon orbits (SPOs) in our study is advantageous for detection purposes. This diversity of unstable SPOs around various rotating black hole models contributes to a more comprehensive understanding and visualization of black holes. In particular, the calculation of the photon ring's two-point correlation function, reliant on unstable SPOs as demonstrated for the Kerr black hole in~\cite{Hadar:2020fda}, holds promise for future validation through experiments like the upcoming Event Horizon Telescope (EHT) project~\cite{Anjum:2023axh}. Secondly, our study sheds light on the crucial information regarding stable spherical timelike orbits (STOs). Being the asymptotic trajectories of homoclinic orbits~\cite{Levin:2008yp, Li:2023bgn}, STOs carry fundamental significance, for example in the context of the gravitational waves analysis from {\it extreme mass-ratio inspirals} (EMRIs)\cite{Teo:2020sey, Ryan:1995zm, Barack:2006pq, Amaro-Seoane:2014ela}. Additionally, our results pave the way for further investigations, such as exploring the gravitational redshift/blueshift emission from STOs around Ghosh black holes, a research avenue previously explored for Kerr and Kerr-de Sitter black holes in \cite{Kraniotis:2019ked}. The other aspect is to explore the black hole merger estimate as presented in \cite{Jai-akson:2017ldo}. It is also interesting to investigate how our results fit into observational constraints by considering the black hole shadow \cite{Tsukamoto:2014tja,Tsukamoto:2017fxq}. We leave this for our further investigations.

%.. Second, perhaps the more important information extracted from this theoretical study is the STOs. As is well-known, STO is the asymptotic trajectory of homoclinic orbits~\cite{Levin:2008yp, Li:2023bgn}, and it is of fundamental importance, for example, in the study of gravitational waves from {\it extreme mass-ratio inspirals} (EMRIs)~\cite{Teo:2020sey, Ryan:1995zm, Barack:2006pq, Amaro-Seoane:2014ela}. Another possible research development from our result is studying the gravitational redshift/blueshift emission from STOs around Ghosh black hole, as was done for KN and KN-de Sitter in~\cite{Kraniotis:2019ked}.  

%===========================================================
\acknowledgments

We thank Edward Teo for the discussion on his analytical spherical orbit solutions. We also thank Leonardo B. Putra for the discussion on ISSO. BNJ is supported by the Second Century Fund (C2F), Chulalongkorn University, Thailand. HSR is funded by the Hibah Riset FMIPA UI No.~PKS-026/UN2.F3.D/PPM.00.02/2023.\\
%===========================================================
%\begin{thebibliography}{9}
\section*{Statement of the Conflict of Interest}

All authors hereby declare that we do not have any conflict of interest.

\section*{Data Availability Statement}

Data sharing is not applicable to this article as no data sets were generated or analyzed during the current study.

%===========================================================

\end{document}